\documentclass[aps,pre,showpacs,twocolumn,notitlepage]{revtex4-1}

\usepackage{bm,amsmath,amssymb,amsfonts,hyperref}

\usepackage{graphicx}
\usepackage{xspace}
\usepackage{color}
\usepackage{layouts}
\usepackage{float}

\allowdisplaybreaks

\newif\ifhyper
\hypertrue
\ifhyper
\hypersetup{
  citecolor = {green},
  urlcolor = {blue} 
} 

\newcommand{\sref}[1]{Sec.~\ref{#1}}
\newcommand{\fref}[1]{Fig.~\ref{#1}}

\newcommand{\aref}[1]{Appendix~\ref{#1}}
\newcommand{\Eq}[1]{Eq.~(\ref{#1})}
\newcommand{\eq}[1]{(\ref{#1})}

\newcommand{\fig}[1]{Fig.~\ref{#1}\xspace}

\newcommand{\ie}{{\it i.e.}\xspace}
\newcommand{\eg}{{\it e.g.}\xspace}

\newcommand{\xx}{\mathbf{x}}
\newcommand{\vx}{\vec{x}}
\newcommand{\vp}{\vec{p}}
\newcommand{\vq}{\vec{q}}

\newcommand{\vk}{\vec{k}}

\newcommand{\pp}{\mathbf{p}}
\newcommand{\qq}{\mathbf{q}}

\newcommand{\GG}{\Gamma}
\newcommand{\vv}{\varphi,\tilde{\varphi}}
\newcommand{\vphi}{\varphi}
\newcommand{\tvphi}{\tilde{\varphi}}
\newcommand{\tr}{\text{Tr}}
\newcommand{\grad}{\vec{\nabla}}
\newcommand{\argf}{-\tilde{D}_t^2,- {\nabla}^2}

\newcommand{\dq}{\frac{\text{d}q}{(2\pi)^d}}
\newcommand{\dom}{\frac{\text{d}\omega}{(2\pi)}}

\newcommand{\p}{\partial}

\newcommand{\fg}{f^{\lambda}_\kappa}
\newcommand{\fl}{f^\lambda_\kappa}
\newcommand{\fn}{f^\nu_\kappa}
\newcommand{\fd}{f^D_\kappa}

\newcommand{\vpi}{\varpi}

\begin{document}

\title{Kardar-Parisi-Zhang Equation with temporally correlated noise: a non-perturbative renormalization group approach}

\author{Davide Squizzato}
\affiliation{Univ. Grenoble Alpes, CNRS, LPMMC, 38000 Grenoble, France}
\author{L\'eonie Canet}
\affiliation{Univ. Grenoble Alpes, CNRS, LPMMC, 38000 Grenoble, France}

\begin{abstract}
We investigate the universal behavior of the Kardar-Parisi-Zhang (KPZ) equation with temporally correlated noise.
The  presence  of time correlations in the microscopic noise breaks the statistical tilt symmetry, or Galilean invariance,
 of the original KPZ equation with delta-correlated noise (denoted SR-KPZ). Thus it is not clear
 whether the KPZ universality class 
 is preserved in this case. Conflicting results exist in the literature, some advocating that it
 is destroyed even in the limit of infinitesimal temporal correlations, while others find that it persists up to a
 critical range of such correlations.  Using non-perturbative and functional renormalization group  techniques, 
 we study the influence of two types of temporal correlators of the noise: a short-range one with a typical time-scale $\tau$, and 
 a power-law one with a varying exponent $\theta$. We show that for the short-range noise with any finite $\tau$,
 the symmetries (the Galilean symmetry, and the time-reversal one in $1+1$ dimension) are dynamically restored at large scales,
 such that the long-distance and long-time properties are governed by the SR-KPZ fixed point. In the presence of a power-law noise, we find that the SR-KPZ fixed point is still stable for $\theta$ below a critical value $\theta_{\textrm{th}}$, in accordance with previous renormalization group results, while a long-range fixed point controls the critical scaling for   $\theta>\theta_{\textrm{th}}$, and
  we evaluate the $\theta$-dependent critical exponents at this long-range fixed point, in both $1+1$ and $2+1$ 
  dimensions. While the results in $1+1$ dimension can be compared with previous studies, no other prediction was available in $2+1$ dimension. 
  We finally report in $1+1$ dimension the emergence of anomalous scaling  in the long-range phase.
\end{abstract}

\maketitle

\section{Introduction}

The Kardar-Parisi-Zhang equation \cite{Kardar86}, originally derived to describe 
 stochastic interface growth,
 stands as a fundamental model in non-equilibrium statistical 
 physics to understand scaling and phase transitions out-of-equilibrium,
  akin the Ising model at equilibrium. 
   Beyond growing interfaces, the KPZ universality class extends to
 many very different systems, such as directed polymers in random media, 
 randomly stirred fluids, particle transport, 
  driven-dissipative Bose-Einstein condensates, to cite a few \cite{Halpin-Healy95,Barabasi95,Krug97,Takeuchi18,Squizzato18}.

An impressive breakthrough has been achieved in the last decade regarding
 the characterization of the KPZ universality class for a one-dimensional interface, 
 sustained by a wealth of exact results \cite{Corwin12}. 
 A particularly striking feature is the discovery of universality sub-classes
  for the distribution of the height fluctuations, determined by the nature of the initial conditions
 (flat, sharp-wedge, or stochastic), which has revealed a 
  deep connection with random matrix theory \cite{Calabrese11,Amir11,Sasamoto10a,Calabrese12,Imamura12}. 
  Moreover, experiments in liquid crystals provided the first set-up to
   allow for quantitative measurements of KPZ universal properties, and 
 they confirmed with a high precision the theoretical results \cite{Takeuchi10,Takeuchi12}.
  
  However, for a higher-dimensional interface, or in the presence of additional ingredients
   such as the presence of correlations of the microscopic noise,
    the integrability of the KPZ equation is broken, and controlled analytical
     methods to describe the rough phase are scarse. The Non-Perturbative (also named functional) Renormalization Group (NPRG) is one of them \cite{Berges02}, 
      and is the one we employ in this work.
      Our aim is to investigate the effect of temporal correlations
       in the microscopic noise on the universal properties of the system. The interest is two-fold. First, strictly uncorrelated processes are a mathematical idealization, any real physical system
 is likely to exhibit some time correlations, at least over a small finite timescale. Hence, it is important to understand their role 
 and assess the relevance of the delta-correlated model. 
  Second, some physical systems are characterized by intrinsic long-range temporal correlations. 
 An interesting example arises in cosmology, where the KPZ equation with power-law time correlations in the noise was shown to emerge
  as an effective model for matter distribution in the Universe, starting from the dynamics of self-gravitating Newtonian fluids \cite{barbero1997,dominguez1999}.
More generally, long-range time correlations can originate from impurities which do not diffuse and impede the growth of the surface, or from the coupling
 of the dynamics to some reservoir which is likely to introduce some memory effects.
      
  Let us now define the model.
 The original KPZ equation describes the stochastic time evolution
  of a height field $h(t,\vx)$, encompassing a smoothening diffusion and a non-linearity as a key ingredient:
\begin{equation}
\label{eq:kpz}
\p_t h=\nu\nabla^2 h+\frac{\lambda}{2}(\grad h)^2 + \eta \, .
\end{equation}
 The non-linear term takes into account a lateral growth of the height profile 
 which tends to enhance the roughening of the interface. The noise $\eta$ is defined
 as a Gaussian noise with zero mean and variance
\begin{equation}
 \langle \eta(t,\vx) \eta(t',\vx\,') \rangle =2 D \, \delta(t-t')\delta^d(\vx-\vx\,')\, ,
\label{eq:delta}
\end{equation}
where $d$ is the dimension of the interface, moving in a $(d+1)$-dimensional space, and $D$ the noise amplitude. 
 As already mentioned, these delta correlations are a simplification, and  this raises the question of the robustness of the 
 KPZ universality class with respect to the presence of some microscopic correlations in the 
  stochastic process driving the growth.
 This question was first investigated by Medina {\it et al.} \cite{Medina89},
  who considered the more general form of noise correlator
\begin{equation}
\langle \eta(t,\vx) \eta(t',\vx\,') \rangle = 2 D(t-t',|\vx-\vx\,'|) \,
\label{eq:etaeta}
\end{equation}
 with long-range (LR) power-law correlations, defined in the Fourier space as 
\begin{equation}
 D(\omega,\vk) = D_0 + D_\theta k^{-2\rho}  \omega^{-2\theta}\, .
\label{eq:powerlaw}
\end{equation}
 This modification of the noise structure breaks the integrability of the original KPZ equation with noise \eq{eq:delta}, that we denote short-range (SR) KPZ. The 
  effect of {\it spatially} correlated noise has been thoroughly
 investigated, both analytically and numerically \cite{Meakin89,Halpin90,Zhang90,Hentschel91,Amar91,Peng91,Pang95,Li97,Chattopadhyay98,Katzav99,Frey99,Janssen99,Verma00,Katzav03,Kloss14a}.  
 It was shown that for a SR enough noise, \ie $\rho<\rho_c$, 
 the standard SR-KPZ properties are preserved, while beyond $\rho_c$, a LR phase
 with $\rho$-dependent critical exponents emerges. For
  a noise characterized by a finite correlation length $\xi$, it was shown  for a one-dimensional interface that the time-reversal symmetry, which is broken
   by the presence of the spatial correlations in the microscopic noise, is restored at large distance, and thus one also finds SR-KPZ universality in this case \cite{Mathey17}.

 In contrast, {\it temporally} correlated noise has received much less attention.
  The few existing analytical
 \cite{Medina89,Ma93,Katzav04,Fedorenko08,Strack15} and numerical \cite{Lam92,Song16,ales2019}
  studies yield conflicting results. One of the reasons is that the presence
   of temporal correlations is much more severe than spatial ones, in that it breaks the constitutive KPZ symmetry,
    which is the Galilean invariance, also known as statistical
    tilt symmetry. Thus it is not clear a priori whether even an infinitesimal
     amount of time-correlation destroys or not KPZ universal physics, and both answers have been given.
   Let us summarize these results.

The problem of temporal correlations of the microscopic noise was first investigated using 
 Dynamical Renormalization Group (DRG) by Medina {\it et al.}, focusing on $d=1$ \cite{Medina89}. They found that the {SR-KPZ} fixed point
 is stable up to a threshold value $\theta_{\textrm{th}}=1/6$, and thus  for $\theta\le \theta_{\textrm{th}}$, the critical exponents are 
the standard SR-KPZ  ones $z_{\rm SR}=3/2$ and $\chi_{\rm SR}=1/2$. Above the threshold $\theta_{\textrm{th}}$,
 they determined from the one-loop flow equations an approximate expression of the critical exponents: 
\begin{equation}
 \chi_{\rm LR} = \frac{1+4\theta}{3+2\theta}\quad,\quad\quad z_{\rm LR} =2-\chi_{\rm LR}\, ,
\label{exp-oneloop}
\end{equation}
obtained by neglecting the corrections on the non-linearity induced by the violation of Galilean invariance due to the temporal correlations. 
This expression is thus only valid for small  $\theta$ close to the threshold.
 Indeed, the exact relation $\chi_{\rm SR}+z_{\rm SR}=2$ stemming from Galilean invariance, in any $d$, only holds at the SR fixed point, 
 and  is replaced at the LR fixed point by the exact relation
\begin{equation}
 z_{\rm LR}(1+2\theta) -2 \chi_{\rm LR} - d=0\, ,
\label{WI-theta}
\end{equation}
which is violated by the estimate \eq{exp-oneloop}. 
The authors then solved numerically a set of truncated  flow equations in $d=1$ which led to exponents, that could be
 approximately fitted by
\begin{equation}
 \chi_{\rm LR} = 1.69 \theta +0.22\quad,\quad\quad z_{\rm LR} =\frac{2\chi_{\rm LR} +1}{1+2\theta}\, .
\label{exp-drg}
\end{equation}
 At variance with this scenario, Ma and Ma \cite{Ma93} advocated on the basis of a Flory-type scaling argument a smooth variation
 of the critical exponents as functions of $\theta$, with no threshold, following
\begin{equation}
 \chi_{\rm LR} = \frac{2+4\theta}{2\theta+d+3}\quad,\quad\quad z_{\rm LR} =\frac{2d+4}{d+3+2\theta}\, ,
\end{equation}
such that the {SR-KPZ} exponents are only recovered at $\theta=0$. This alternative scenario was supported 
by a Self-Consistent Expansion (SCE) developed by Katzav and Schwartz \cite{Katzav04}. The authors found within the SCE two strong-coupling  solutions, one which coincides with the one-loop  DRG result,
 and the other, considered as dominant, which leads to a smooth dependence on $\theta$ with no threshold, and with a decreasing $z_{\rm LR}(\theta)$,
 whereas the solution \eq{exp-drg} is increasing.

 The problem was re-visited using perturbative functional RG within the framework of  elastic manifolds in correlated disorder \cite{Fedorenko08}.
In this context, a crossover from a SR behavior to a LR one beyond a certain threshold was confirmed.
 The two-loop LR exponents were calculated in a perturbative expansion in $\epsilon=4-d$ where $d$
 is the dimension of the elastic manifold. However, the KPZ interface is equivalent to a $d=1$ directed polymer,
 which implies $\epsilon=3$, and the extrapolation to such a large value is not reliable. Notwithstanding this 
 limitation, the two-loop results indicate a decreasing $z_{\rm LR}$ for small $\theta$, at variance with \eq{exp-drg}.  
 Based on a stability criterion, the author also derives bounds for the value of $z_{\rm LR}$ in $d=1$ as
\begin{equation}\label{eq:fedbounds}
 \frac{5}{3+2\theta}\le z_{\rm LR} \le \frac 3 2
\end{equation}
where the lower bound coincides with the one-loop result \eq{exp-oneloop}.
 This bound rules out both the second  SCE solution and the scaling solution.
 On the analytical side, the situation is thus unclear.

On the numerical side, very few attempts exist in the literature. Among them,
 Refs. \cite{Lam92,Song16} cannot convincingly discriminate between the two scenarii (presence or absence of a threshold)
 nor on the sense of variation of  $z_{\rm LR}$. 
  They essentially find a very weak dependence at small $\theta$ and are too scattered to settle whether $z_{\rm LR}$ is decreasing or increasing at larger values of $\theta$.
A progress in this direction was recently achieved in Ref. \cite{ales2019}, where the authors simulate both the KPZ equation and ballistic deposition with temporal correlations with improved accuracy. They find no threshold, that is the appearance of a long-range phase for any non-zero temporal correlation. Moreover, they unveil the existence of anomalous scaling
 at large $\theta$, which they relate to the emergence of ``faceting'' structures \cite{ales2019}.

 Note that the effect of temporal correlations is also crucial in the context of turbulence.
 In particular,  field theoretical approaches to turbulence are constructed from  Navier-Stokes
 equation with a stochastic large-scale forcing, which is delta-correlated in time to preserve Galilean invariance,
  whereas a physical forcing cannot be completely uncorrelated.
 The presence of temporal correlations in the forcing correlator was investigated 
 in  \cite{Antonov18}, and the results support the robustness of the SR properties below a threshold value.

In this work, we analyze the effect of temporal correlations in the microscopic noise of the KPZ equation
 in the framework of the 
 NPRG. Indeed, this method has turned out to be successful to describe KPZ interfaces
 since the NPRG flow equations embed  the strong-coupling fixed point in any dimensions
    \cite{Canet10},  whereas the latter cannot be reached at any order from perturbative expansions \cite{Wiese97}. 
 Moreover, a controlled approximation scheme, based on symmetries, can be devised in this framework \cite{Canet11a,Kloss12}. 
  It was shown that it reproduces with very high accuracy the exact results in $d=1$ for the scaling function \cite{Canet11a}.
 It yielded predictions for dimensionless ratios in $d=2$ and $3$ \cite{Kloss12} which were later accurately confirmed by large-scale numerical simulations \cite{Halpin-Healy13,*Halpin-Healy13Err}. This framework was extended to study the effect of anisotropy \cite{Kloss14b}, and also of spatial correlations in the noise,  following a power-law \cite{Kloss14a} or with a finite length-scale \cite{Mathey17}. 

We here study the influence of temporal correlations both in $d=1$ and 
$d=2$, and both for a finite correlation time or for a LR power-law correlator
 \begin{equation}
 D_\tau(\omega,\vk) =D_0 e^{-\frac{1}{2}\omega^2\tau^2},\quad \quad D_\infty(\omega,\vk) = D_0+D_\theta\omega^{-2\theta}\, .
\label{eq:choiseD}
\end{equation}
 The $D_\tau$ correlator is studied to probe whether the {SR-KPZ} physics is destroyed
 as soon as Galilean invariance is broken at the microscopic scale, even on a short finite range. We find that this is not the case, and we show that when $\tau$ is finite,
 this symmetry is always restored at long distance and long time.
 We then investigate the effect of the power-law temporal noise $D_\infty$, and find that the {SR-KPZ} fixed point is stable
 below a threshold ${\theta_{\textrm{th}}}=1/6$ in $d=1$ and ${\theta_{\textrm{th}}}\simeq 0.35$ in $d=2$. Beyond this threshold, a LR fixed-point takes
 over and we compute the $\theta$-dependent critical exponents in this LR dominated phase. 
 We find  that $z_{\rm LR}(\theta)$ is decreasing and satisfy the bound \eq{eq:fedbounds} in $d=1$.
We finally investigate in more details the scaling properties of the LR phase in $d=1$,
 which shows the presence of anomalous scaling, in agreement with the results from the numerical simulations of \cite{ales2019}.

 The remainder of the paper is organized as follows. We briefly present the KPZ field theory and its symmetries in \sref{sec:KPZ}.
 We then introduce the NPRG framework, and the approximation scheme used in \sref{sec:NPRG}, and derive the corresponding flow equations.
 The results are presented and discussed in \sref{sec:RES}.

\section{KPZ field theory and its symmetries}
\label{sec:KPZ}

The KPZ equation \eq{eq:kpz} can be cast into a field theory following the standard response functional formalism introduced by Martin-Siggia-Rose and Janssen-De Dominicis \cite{Martin73,Janssen76,Dominicis76}. The KPZ field theory reads
\begin{align}
{\cal Z}[j,\tilde j] =& \int {\cal D}[h]{\cal D}[\tilde h] e^{-{\cal S}[h,\tilde h]+\int_{t,\vx} \big\{ jh+\tilde j\tilde h\big\}}\nonumber\\
{\cal S}[h,\tilde{h}]=&\int_{t,\vx}\left\{ \tilde{h}\left( \partial_t h- \nu \nabla^2 h-\frac{\lambda}{2}(\grad h)^2 \right)\right\}\nonumber\\
-&\int_{\omega,\vq} \tilde{h}(-\omega,-\vq) D(\omega,\vq) \tilde{h}(\omega,\vq)
\label{eq:KPZaction}
\end{align}
where $\int_{t,\vx}\equiv \int dt d^d\vx$ and  $\int_{\omega,\vq}\equiv \int\frac{d\omega}{2\pi}\frac{d^d\vq}{(2\pi)^{d}}$.
Upon rescaling the time and the fields, one finds that the SR part of the KPZ action is characterized by a single dimensionless
 coupling  $g=\lambda^2 D_0/\nu^3$, while  the LR correlation introduces another dimensionless coupling
  $w_\theta =  D_\theta/(D_0\nu^{2\theta})$. These couplings have canonical dimensions
\begin{equation}
[g]=2-d, \quad [w_\theta]=4\theta\, .
\end{equation}
In the absence of temporal correlations, \ie with a noise correlator $D_0(\omega,\vk)=D_0$, the KPZ action possesses several symmetries.
 Besides the usual invariance under space-time translations and space rotations, 
 it  is invariant under a shift in the height field and a Galilean transformation (or tilt of the interface).
 The latter enforces the exact relation
$z+\chi=2$ in any dimension.

In fact, theses last symmetries admit extended forms, which  correspond to the following  infinitesimal field transformations
 with time-dependent parameters:
\begin{equation}
h(t,\vx) \longrightarrow h(t,\vx)+ c(t)
\label{shift}
\end{equation} 
for the height shift, and for the Galilean transformation
\begin{align}
h(t,\vx) &\longrightarrow h(t,\vx+\lambda \vec{\epsilon}(t))+ \vx \cdot \p_t \vec{\epsilon} \nonumber \\
\tilde{h}(t,\vx) &\longrightarrow \tilde{h}(t,\vx+\lambda \vec{\epsilon}(t))\, .
\label{gal}
\end{align}
The choice $\vec{\epsilon}(t) = \vec{v} \, t$ yields the standard Galilean transformation
 (for the velocity field $\grad h$, which corresponds to a tilt for the height field).  An arbitrary infinitesimal $\vec{\epsilon}(t)$ gives a local-in-time, or time-gauged
 Galilean transformation. The time-gauged symmetries \eq{shift} and \eq{gal} are extended symmetries, in the sense that the KPZ action is not
 strictly invariant under these transformations, but the induced variations are linear in the fields. 
 One can also derive in the case of extended symmetries  Ward identities which, because of the locality in time of the corresponding transformations,  have a stronger content than their non-gauged versions \cite{Canet11a}.
 These exact identities are very useful to constrain approximations.
 For a $d=1$ interface, there exists an additional discrete symmetry associated to the time-reversal transformation \cite{Canet05}
\begin{align}
h(t,\vx) &\rightarrow -h(-t,\vx) \nonumber \\
\tilde{h}(t,\vx) &\rightarrow \tilde{h}(-t,\vx) + \frac{\nu}{D_0} \nabla^2 h(-t,\vx) \,.
\end{align}
Indeed the corresponding variation of the action  is $\delta {\cal S} \propto \int_{\vx} (\grad h)^2 \nabla^2 h$,
  which vanishes in one dimension only. The existence of this additional symmetry in $d=1$ in turn completely fixes the SR-KPZ critical exponents in this dimension
 to the values $\chi_{\rm SR}=1/2$ and $z_{\rm SR}=3/2$.

 The presence of temporal correlations, either of the form $D_\tau$ or $D_\infty$, breaks the Galilean symmetry
 in all dimensions, and also the time-reversal symmetry in $d=1$. The consequences are studied within the NPRG, which is presented in the next section.

\section{Non-Perturbative Renormalization Group for KPZ}
\label{sec:NPRG}

\subsection{Non-Perturbative Renormalization Group formalism}

 Integrating out microscopic fluctuations plays a central role in understanding the long-distance and long-time
  universal properties of a physical system.
The NPRG is a modern implementation of
 Wilson's original idea of the renormalization group \citep{Wilson74}, conceived to efficiently average over fluctuations, even when they
 develop at all scales, as in standard critical phenomena \citep{Berges02,Kopietz10,Delamotte12}. It is a powerful 
 method to compute the properties of strongly
  correlated systems,  which can reach high precision levels \citep{Canet03b,Benitez12,Balog19},
  and can yield fully non-perturbative results, at equilibrium \citep{Grater95,Tissier06,Essafi11} and also 
  for non-equilibrium systems \citep{Canet04a,Canet05,Canet10,Canet11a,Berges12,Tarpin17}, restricting to a few classical statistical
   physics applications. 

  The progressive integration of fluctuation modes is achieved by introducing in the KPZ action \eq{eq:KPZaction} a scale-dependent quadratic term
\begin{equation}
\Delta {\cal S}_\kappa=\frac{1}{2}\int_{\omega,\vq} \phi_i(\omega,\vq) [R_\kappa (\omega,\vq)]_{i,j} \phi_j(-\omega,-\vq)
\end{equation}
where $\kappa$ is a momentum scale, and $\phi_1 \equiv h$, $\phi_2\equiv \tilde h$.
The matrix elements of $R_\kappa$ are proportional to a cutoff function
 $r(q^2/\kappa^2)$, with $q=|\vq|$, which ensures the selection of 
fluctuation modes: $r(x)$ is required to be large  for  $x\lesssim 1 $ such that 
the fluctuation modes $\phi_i(q \lesssim \kappa)$ are essentially frozen and do not contribute
 in the path integral, and to be negligible  for $x\gtrsim 1 $ such that the other modes ($\phi_i(q\gtrsim \kappa)$)
 are not affected.
 $\Delta {\cal S}_\kappa$ must preserve the symmetries of the original action and causality properties.
For the KPZ field theory, a suitable form is \cite{Canet10}
 \begin{equation}
R_\kappa(\omega,\vq) \!\equiv\!R_\kappa(\vq) \!=\! r\left(\frac{q^2}{\kappa^2}\right)
\left(\!\! \begin{array}{cc}
0& {\nu_\kappa} q^2\\
{\nu_\kappa} q^2 & -2 D_\kappa 
\end{array}\!\!\right) \;,
\label{Rk}
\end{equation}
where the running coefficients $\nu_\kappa$ and $D_\kappa$ are defined later.
Here we work with the cutoff function 
\begin{equation}
r(x)=\alpha/(\exp(x) -1)\, ,
\label{eq:expReg}
\end{equation}
where $\alpha$ is a free parameter. In the exact theory, the results are independent of the precise form of the cut-off function.
 However, any approximation introduces a (typically small) spurious dependence on this choice. The parameter $\alpha$
  can thus be conveniently used to estimate the error and optimize the results, as discussed in \aref{sec:cutoff}.

We emphasize that the regulator $\eq{Rk}$ does not depend on frequency. Whereas it would be desirable
 to also regularize in frequency, it is  much simpler  not to, and it is the actual choice
  made in most applications to non-equilibrium systems \cite{Canet04a,Canet11b}. It turns out that 
  for most applications, regularizing in momentum is enough to achieve the separation of fluctuation modes and to ensure the analyticity of the flow. 
  The implementation of a frequency regularization
   was studied in \cite{Duclut17} on the example of Model A, where it was shown that it does improve the results.
    However, the difficulty lies in formulating a regulator which respects both causality and all the symmetries
    of the model.  For KPZ, the Galilean invariance precludes from having a (manageable) frequency-dependent regulator.
  This has implications for the study of the power-law correlator $D_\infty$, since the latter brings non-analyticities
  in $\omega$  which would be cured (as they should) by a frequency regularization, whereas with only a momentum regulator they can 
 survive and have to be dealt with (as explained in the following).

The inclusion of $\Delta {\cal S}_\kappa$ in \eq{eq:KPZaction} leads to a scale-dependent generating functional ${\cal Z}_\kappa$.
 Field expectation values in the presence of the external sources $j$ and $\tilde{j}$ are obtained from the functional
   ${\cal W}_\kappa = \log {\cal Z}_\kappa$ as 
\begin{equation}
  \varphi({\bf x}) = \langle h({\bf x}) \rangle = \frac{\delta {\cal W}_{\kappa}}{\delta j({\bf x})}  \, \, , \, \,
  \tilde \varphi({\bf x}) = \langle \tilde h({\bf x}) \rangle = \frac{\delta {\cal W}_{\kappa}}{\delta \tilde j({\bf x})}  \, ,
\end{equation} 
denoting ${\bf x} = (t,\vx)$.
 The effective average action  is defined as the modified Legendre transform of ${\cal W}_\kappa$ as
 \begin{equation}
  \Gamma_\kappa[\varphi,\tilde\varphi] +{\cal W}_\kappa[j,\tilde j] = 
\int \! j_i \varphi_i -\frac{1}{2} \int \varphi_i \, [R_\kappa  ]_{ij}\, \varphi_{j} .
\label{legendre}
\end{equation}
where $j_i$ are the sources associated with the fields $\varphi_i$, with $\varphi_1=\varphi$, $\varphi_2=\tilde \varphi$ and similarly for $j_i$. 
 The last term in (\ref{legendre}) ensures that at the microscopic scale $\Lambda$, the effective average action coincides with the microscopic
 action $\Gamma_{\kappa=\Lambda}={\cal S}$,	 provided that $R_\kappa$ is very large when $\kappa\to \Lambda$ \cite{Berges02}. In the opposite limit $\kappa\to0$, 
 the cut-off $R_\kappa$ is required to vanish such that one recovers
 the standard effective action $\Gamma$ (which would be Gibbs free energy for an equilibrium system). The scale-dependent effective average action $\Gamma_\kappa$ thus smoothly interpolates between the microscopic action
 and the full effective action which encompasses  all the fluctuations.
 It obeys an exact flow equation, usually referred to as Wetterich equation \cite{Wetterich93}:
\begin{equation}\label{eq:weteq}
\partial_s \Gamma_\kappa = \frac{1}{2} \tr \left\{ \partial_s R_\kappa \, \left[ \Gamma_\kappa^{(2)} + R_\kappa \right]^{-1} \right\}
\end{equation}
where $s=\ln(\kappa/\Lambda)$ is the renormalization ``time'' and $\Gamma_\kappa^{(2)}$ is the $2\times 2$ Hessian matrix
\begin{equation}
[\Gamma^{(2)}_\kappa]_{i,j}=\frac{\delta^2 \Gamma_\kappa[\{\varphi\}]}{\delta \varphi_i \, \delta \varphi_j} \,.
\end{equation}
$\tr\{\cdot\}$ is the trace over all the internal degrees of freedom.

Eventhough the equation \eqref{eq:weteq} is exact, it cannot be solved exactly because of its non-linear functional integro-differential structure.
One has to employ some approximation scheme \cite{Berges02}. The key advantage of this approach is that these approximations 
 do not have to be perturbative in couplings or in dimensions, but they are rather based on some controlled truncation
  of the functional space. 
  There exist two main approximation schemes within the NPRG context: the derivative expansion  \cite{Berges02} 
 and the  Blaizot-Mendez-Wschebor (BMW) scheme \cite{Blaizot06,Benitez09}. The derivative expansion, which is the most widely used, consists in
 expanding the effective average action $\Gamma_\kappa$ in powers of gradients and time derivatives, retaining a finite number of terms. It usually provides
 a reliable description of large-distance and long-time properties (that is the small momentum and frequency sector), including critical exponents and phase diagrams.
 Furthermore, it can reach a  high precision
 level, competing with current boostrap methods for the Ising model \cite{Balog19}. 
 On the other hand, the BMW scheme is designed to 
 reliably obtain the full momentum and frequency dependence of the correlation functions, not limited to the small momentum and frequency sector.
 It rather relies on an expansion 
 in the vertices of the flow equations, which is controlled by the presence of the regulator term.  
  It was also shown to reach a high precision \cite{Benitez12}.

  For the KPZ equation, the simplest approximation is the first order of the derivative expansion, which is usually called
 the Local Potential Approximation (LPA). This approximation was shown to be sufficient to access the strong-coupling KPZ fixed point in any dimensions $d$ \cite{Canet05b}.
  It thus already goes beyond perturbative RG, since the latter fails to capture this fixed point in $d\neq 1$ even to all orders in perturbation theory \cite{Wiese97}.
  However, the critical exponents are quite poorly determined within LPA, except in $d=1$ where they are fixed exactly by the  symmetries.
 This lack of accuracy of the derivative expansion for the KPZ problem is related to the 
 derivative nature of the interaction in the KPZ equation. The BMW scheme
 has turned out to be more appropriate in this context, as evidenced in subsequent studies \cite{Canet11a,Kloss12}. In fact,
 the standard BMW scheme has to be adapted in order not 
 to spoil the KPZ symmetries. Its rationale is expounded in more details in \cite{Canet11a,Kloss12}. In practice, it can be implemented
    using an ansatz for the effective average action, which is presented in the next section. 

\subsection{Effective average action for the pure {SR-KPZ}}

To study the original KPZ equation, one can use an ansatz for $\Gamma_\kappa$, such that i) it preserves the
 full momentum and frequency dependence of the two-point functions, and ii) it preserves the KPZ symmetries.
 Using an ansatz, rather than performing a direct BMW expansion of the vertices, is 
  a solution to concile i) and ii). Indeed, the (extended) Galilean symmetry yields constraints on the vertices  $\Gamma_\kappa^{(n)}$, under the form of exact Ward identities, which relate 
 a $\Gamma_\kappa^{(n+1)}$ vertex with one vanishing momentum on a $\varphi$ leg to a lower order vertex $\Gamma_\kappa^{(n)}$.
 Introducing the notation $\Gamma_\kappa^{(m,n)}$ where the $m$ first derivatives are with respect
 to $\varphi$ and the $n$ last with respect to  $\tilde \varphi$, they read  \cite{Canet11a}:
\begin{align}
  &\frac{\p}{\p \vq}\Gamma_\kappa^{(m+1,n)}(\omega,\vq,\varpi_1,\vp_1, \dots , \varpi_{n+m-1},\vp_{n+m-1})\Big|_{\vq=0}\nonumber\\
 &= -i\lambda \sum_{k=1}^{n+m-1}\frac{\vp_k}{\omega} \Big[\Gamma_\kappa^{(m,n)}(\cdots,\varpi_k+\omega,\vp_k,\cdots) \nonumber\\
  &\quad\quad\quad\quad - \Gamma_\kappa^{(m,n)}(\cdots ,\varpi_k,\vp_k,\cdots)  \Big]
 \, . \label{eq:wardGalN}
\end{align}
 Expanding the vertices while satisfying these identities turns out to be complicated.
 A simpler way is to construct
 a general ansatz for the effective average action 
 using as building blocks invariants under the Galilean symmetry. For this symmetry,
 one can define a field $f(t,\vx)$ as a scalar density if its infinitesimal transform under \eq{gal}
 is $\delta f(t,\vx) = \lambda \vec \epsilon(t)\cdot\grad f$, since this implies that $\int d^d \vx f$ is invariant under a Galilean transformation.
 One can check that with this definition, 
 the elementary  Galilean scalar densities are $\tilde h$, $\p_i \p_j h$, and 
\begin{equation}
  D_t h \equiv \p_t h-\frac{\lambda}{2}(\grad h)^2\, ,
\end{equation}
 but not $\p_t h$ alone.
 The scalar property  is preserved by the operator $\grad$ and by the covariant time derivative
\begin{equation}  
\tilde D_t = \p_t- \lambda\grad h \cdot \grad\, ,
\end{equation}
but not by $\p_t$.
Combining these Galilean scalars and operators, one can  construct an ansatz which explicitly preserves Galilean symmetry.  
At quadratic order in the response field, the most general ansatz obtained in this way, called SO (for Second Order),
  was first proposed in \cite{Canet11a} and reads:
\begin{align}
\Gamma_\kappa&[\vv]=\int_\xx \left\{ \tvphi \fl(\argf)D_t\vphi- \tvphi \fd(\argf)\tvphi \right.&\nonumber\\
 & \left.-\frac{1}{2}\Big[\nabla^2 \vphi \fn(\argf) \tvphi  + \tvphi \fn(\argf)\nabla^2\vphi\Big] \right\} 
\label{eq:anzNLO}
\end{align}
with $f_\kappa^X$ analytic functions of their arguments defined as
\begin{equation}
f^X_\kappa(-\tilde{D}_t^2,-\nabla^2)=\sum_{m,n=0}^\infty a_{\kappa,mn}^X (-\tilde{D}_t^2)^m (-\nabla^2)^n \, .
\end{equation}
One notices that the  term proportional to $D_t\varphi$ renormalizes as a whole, with a unique function $f_\kappa^\lambda$ in \eq{eq:anzNLO},
 which is equivalent to stating that $\lambda$ is not renormalized.
The Ward  identities \eq{eq:wardGalN} are automatically satisfied at all scales $\kappa$ by the vertices $\Gamma_\kappa^{(n)}$ computed from the ansatz \eq{eq:anzNLO} \cite{Canet11a}.
Furthermore, additional constraints stem from the other symmetries. 
The time-gauged shift symmetry \eq{shift} imposes that $\fl(\omega,\vp=0)=1$ at any scale $\kappa$. 
 In $d=1$, the time-reversal symmetry further imposes that $\fd = \fn$, and  $\fl =1$, such that there is a single independent running function in one dimension.

The ansatz \eq{eq:anzNLO} truncates the functional dependence in $\tilde \varphi$ at quadratic order, but
 it remains functional in  $\varphi$ through  the operators $\tilde{D}_t$. 
 This ansatz provides a non-trivial frequency and momentum dependence for all vertices $\Gamma_\kappa^{(n)}$. This dependence is the most general one  for the two-point functions, but it is not for higher order vertices.
 It was shown in \cite{Canet11a} that this ansatz yields very accurate results. It reproduces in particular to a very high precision level
  the exact results available in $d=1$ for the scaling functions associated with the two-point correlation function, up to very fine details of the tails of these functions.

 However, solving the flow equations at SO represents quite a heavy numerical task in $d>1$.
 Thus, a simplification was proposed in \cite{Kloss12}, which consists in neglecting the frequency dependence of the functions $f_\kappa^X$ {\it within  the integrands}
  of the flow equations.  
  This approximation, named Next-to-Leading Order (NLO), allows one to explore higher spatial dimensions in a reasonable computational time.
  Indeed, at NLO, all the $n-$point vertices $\Gamma^{(n)}_\kappa$, with $n>2$, vanish except the bare one $\Gamma^{(2,1)}_\kappa$.
  The NLO approximation leads to reliable estimates for the critical exponents in $d=2$ and $d=3$, 
   and it enables one to determine non-trivial properties of the rough phase, such as scaling functions,
    and associated universal amplitude ratios \cite{Kloss12}. Some of the predictions obtained at NLO were accurately confirmed by subsequent numerical simulations \cite{Halpin-Healy13,*Halpin-Healy13Err}.
    Note that this approximation turns out to deteriorate when the dimension grows, and it becomes unreliable
     above $d\gtrsim 3.5$ \cite{Kloss12}. Therefore it cannot be used for instance to probe the existence or not of an upper critical
      dimension for KPZ, for which the full SO approximation should be implemented.
      
      In this work, we use approximations close to the NLO one,  minimally extended to take into account violations of Galilean invariance. 
 For the LR noise, we simply
       include the scale-dependent long-range coupling constant
 $w_\kappa^\theta$ associated with the non-analytical frequency dependence of the effective noise, together with the induced renormalization of the non-linear coupling $\lambda_\kappa$.
 These two quantities are enough to discriminate between a LR and a SR phase and to estimate the corresponding critical exponents, as shown in the following.
 In this case, the analytical frequency dependence, carried by the renormalization functions $f_\kappa^X(\omega,p)$, is sub-dominant compared to the non-analytical one,
 so it is sufficient to compute it within the NLO approximation. 
 For the SR noise, all the non-trivial frequency dependence generated by this noise is carried by the analytical function $f^D_\kappa$. In particular, at the microscopic scale 
 $\Lambda$, this function takes the form $D_\tau$ in \eq{eq:choiseD}. To implement this initial condition, one cannot neglect
 the frequency dependence of  $f_\kappa^D$ in the right-hand side of the flow equations, as is done in NLO. Hence we devised
   a generalized version, denoted NLO$_\omega$,  which keeps the  frequency dependence of the functions $f^\nu_\kappa$ and $f^D_\kappa$ in the integrands of hte flow equations. 

Since the NLO
  can be obtained as a simplification of the  NLO$_\omega$, we present first  in the next section the  latter approximation, 
  and then the subsequent simplifications. These different approximations, LPA, NLO, NLO$_\omega$ and SO, can be seen as four successive orders of our approximation scheme, with increasing accuracy.
  We emphasize that the NLO order is already a satisfactory (and quite involved) one since a good accuracy can be obtained at this order in the physical dimensions 1, 2 and 3 \cite{Kloss12}. 

\subsection{Effective average action with broken Galilean invariance}

  Introducing a non-trivial frequency dependence in the  noise correlator of the KPZ equation breaks  Galilean invariance at the microscopic level.
 This means that the constraints associated with this symmetry no longer apply.
 In particular, the non-linear coupling  can acquire a non-trivial RG flow  $\lambda \equiv \lambda_k$, since it is no longer the structure constant of
 a symmetry  of the system. This renormalization has to be taken into account. 
 It implies in particular that  the covariant time derivative is splitted in two independent parts.
 This induces some modifications of the ansatz. First, the term proportional to $D_t\varphi$
  separates in two parts
\begin{equation}
\fl \, D_t\vphi \rightarrow \fl \, \p_t \vphi -\frac{\lambda_\kappa}{2} \fl \, (\grad \vphi)^2\, ,
\end{equation}
and we keep, as a minimal extension of the NLO approximation, the same function $\fl$ for the two parts. Although they can in principles be different,
 it is clear that the dominant effect of the breaking of Galilean symmetry is described by the renormalization of $\lambda_\kappa$.
Similarly, $\tilde D_t$ decomposes in two independent parts. For simplicity, again as a minimal extension of NLO,
 we only retain at NLO$_\omega$ in the arguments of  the functions $f_\kappa^X$ the time derivative part when necessary, that is
  for $f^D_\kappa$ and  $f^\nu_\kappa$
\begin{equation}
f^{D,\nu}_\kappa (-\tilde{D}_t^2,-\nabla^2) \rightarrow f^{D,\nu}_\kappa (-\p_t^2,-\nabla^2)\, .
\end{equation} 
For $\fl$, better resolving its frequency dependence is not needed, so we compute it only in the NLO approximation.
Thus, within the NLO$_\omega$ approximation,  the functions $f_\kappa^X$ no longer depend on the field $\varphi$, 
which implies that only the 3-point vertex $\Gamma_\kappa^{(2,1)}$ is non-zero, as for NLO. 
The corresponding ansatz NLO$_\omega$, reads
\begin{align}\label{eq:nlopansatz}
\Gamma_\kappa[\vv]&=\int_\xx \left\{ \tvphi \fl \p_t\vphi-  \frac{\lambda_\kappa}{2}\tvphi \fg (\grad\vphi)^2 -\tvphi \fd\tvphi\right.\nonumber\\
& \left. -\frac{1}{2}\left[\nabla^2 \vphi \fn\tvphi + \tvphi \fn\nabla^2\vphi\right] \right\}\, ,
\end{align}
where all functions depend on $(-\p_t^2,-\nabla^2)$.
 With this ansatz, the two-point functions are given by
\begin{align}
 \Gamma_\kappa^{(1,1)}(\varpi,\vp) & = i \varpi \fl(\varpi,p) + \vp^{\, 2}  \fn(\varpi,p)\nonumber \\
  \Gamma_\kappa^{(0,2)}(\varpi,\vp) & = -2  \fd(\varpi,p)\nonumber\\
 \Gamma_\kappa^{(2,0)}(\varpi,\vp) & =0\, ,
\label{eq:2point}
\end{align}
noting simply  that the actual dependence of the functions $f_\kappa^X$ is on $\varpi^2$ and $\vp^{\, 2}$,
 and that the frequency dependence of $\fl$ is treated in the NLO approximation, {\it i.e.} it never appears
  in the non-linear part  of the flow equations \eq{eqf}.

Let us place again this approximation with respect to the other ones, NLO and SO.
In the NLO approximation, the frequency dependence of all the functions $f^X_\kappa$ is neglected in the right-hand side of the flow equations, which amounts to the replacement $f_\kappa^X(\varpi,p)\to f_\kappa^X(p)$ in the integrands of \eq{eqf} \cite{Kloss12}.
 The functions $f^X_\kappa$  nonetheless acquire a frequency dependence, which is generated by the explicit  dependence on the external frequency  in the flow equations (through $P_\kappa(\Omega,Q)$ in \eq{eqf}). 
 Within the NLO$_\omega$ approximation, this replacement is performed only for the function $f_\kappa^\lambda$, while the full frequency dependence of $f_\kappa^D$ and $f_\kappa^\nu$ is kept
 in the flow equations. This is the minimal approximation that allows one to study SR temporal correlations in the noise while limiting the explicit breaking of the KPZ symmetries by the ansatz. The NLO$_\omega$ scheme induces an additional computational cost compared to NLO (in particular, the integration over the internal frequency $\omega$ can no longer be performed analytically, and additional interpolations in the frequency sector are needed, see \aref{sec:num}).
 However, since the NLO$_\omega$ approximation is actually quadratic in both $\varphi$ and $\tilde\varphi$, there remains only 
one  non-zero  3-point vertex function, as in the NLO scheme, which is $\Gamma_\kappa^{(2,1)}$
\begin{align}
\label{gamma21}
\Gamma^{(2,1)}_{\kappa}(\varpi_1,\vp_1,\varpi_2,\vp_2)=\lambda_\kappa\vp_1\cdot\vp_2  \fg((\varpi_1+\varpi_2)^2,|\vp_1+\vp_2|^2).
\end{align}
 This implies that the expression of the flow equations is still greatly simplified compared to the SO scheme \cite{Canet11b}, and thus remains
 numerically reasonable, in particular in $d=2$. The price to pay is that the NLO$_\omega$ approximation induces a small spurious breaking of the Galilean invariance (even when this symmetry is present at the microscopic level). Indeed, contrarily to the $\tilde{D}_t$ operator, the simple time derivative $\p_t$ does not preserve the Galilean scalar property. In particular, the frequency dependence in the two-point functions is not accompanied by a higher-order field dependence as it should to satisfy the Galilean Ward identities (\ref{eq:wardGalN}) and thus preserve this symmetry. Treating the full frequency dependence without inducing any spurious breaking of Galilean invariance would require to work with the SO ansatz. However, within the NLO$_\omega$ scheme, this spurious breaking remains very small, and does not prevent from identifying a ``true'' physical breaking, as shown in the next sections.
  In fact, it provides an estimate of the error associated with this order of approximation, which is small. 

\begin{figure*}[ht]
\includegraphics[scale=0.8]{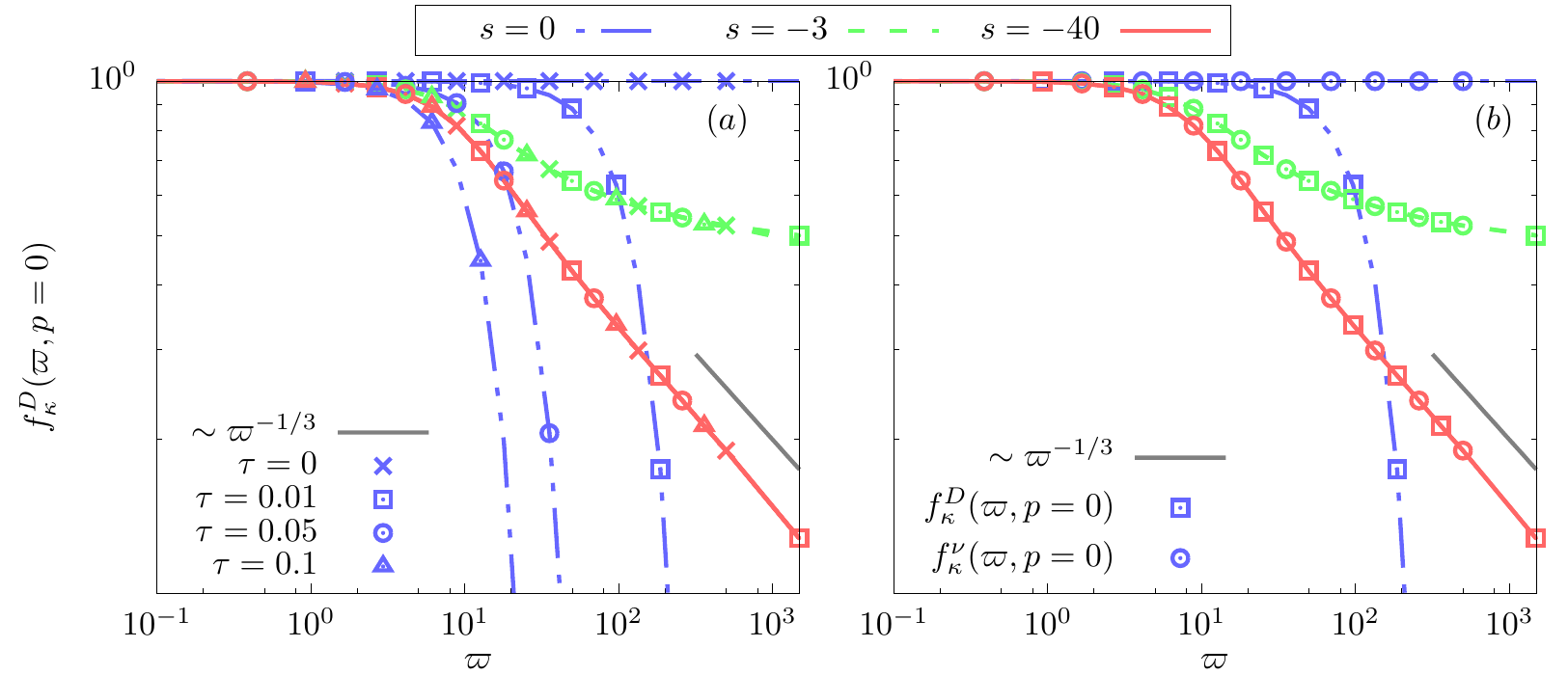}
\caption{(a) Evolution of the function $f^D_{\kappa}(\varpi,0)$ with the RG scale for different values of $\tau=0,0.01,0.05,0.1$. For each $\tau$, the function is represented at successive RG times $s= -\log(\kappa/\Lambda)$:
 $s=0$ where they are Gaussians of different width, $s=3$ where the functions for the different $\tau$ are already 
 almost superimposed (erasure of the initial conditions), and $s=40$ where the fixed-point shape is reached,  characterized 
 by a power-law decay very close to the KPZ one $\sim \varpi^{1/3}$ (indicated as a guideline). (b) Evolution of the functions $f^D_{\kappa}(\varpi,0)$ and $f^\nu_{\kappa}(\varpi,0)$ with the RG scale for $\tau=0.01$, represented for the  RG times $s=0,3,40$. Although they start at $s=0$ from different shapes $f^D_{\kappa=\Lambda}(\varpi,0)\neq f^\nu_{\kappa=\Lambda}(\varpi,0)$, the time-reversal symmetry is restored at $s\lesssim 3$ where they already coincide, up to the fixed point $f^D_{*}(\varpi,0)=f^\nu_{*}(\varpi,0)$.}
\label{fig:fomSRd1}
\end{figure*}

\subsection{Flow equations and running anomalous dimensions}
\label{sec:flow}

\subsubsection{Flow equations of the renormalization functions $f_\kappa^X$}
According to \eq{eq:2point}, the flow equation for the running functions $\fn$, $\fl$, and $\fd$ respectively, can be deduced from the flow
 equations of the real part, imaginary part of  $\Gamma_\kappa^{(1,1)}$, and  $\Gamma_\kappa^{(0,2)}$, respectively. 
 One has to take two functional derivatives of the exact flow equation (\ref{eq:weteq}), and then replace the vertex functions
 and the propagator in this expression by the ones computed from the ansatz (\ref{eq:nlopansatz}), evaluated at zero fields.
 The calculations are the same as those reported in \cite{Kloss12}, where more details can be found. One obtains within the NLO$_\omega$ scheme
\begin{widetext}
\begin{subequations}
\begin{align}
\p_\kappa f_\kappa^D(\varpi,p) & = 2 g_\kappa f_\kappa^\lambda(p)^2\displaystyle \int_{\omega,\vq} \frac{(\vq\,^2 +(\vp\cdot\vq))^2 \,k_\kappa(\Omega,Q)}{P_\kappa(\omega,q)^2 P_\kappa(\Omega,Q)}\Bigg\{P_\kappa(\omega,q)\, \p_\kappa S_\kappa^D(q)-2\, \vq\,^2 \,\ell_\kappa(\omega,q)\,k_\kappa(\omega,q)\, \p_\kappa S_\kappa^\nu(q) \Bigg\} , \label{dsfd}\\
 \p_\kappa f_\kappa^\nu(\varpi,p) & =\displaystyle -2\frac{g_\kappa}{p^2}  f_\kappa^\lambda(p)\int_{\omega,\vq}  \frac{\vq\,^2 +(\vp\cdot\vq)}{P_\kappa(\omega,q)^2 P_\kappa(\Omega,Q)}\Bigg\{-\vp\cdot\vq \, f_\kappa^\lambda(Q)\, \ell_\kappa(\Omega,Q)\, P_\kappa(\omega,q) \, \p_\kappa S_\kappa^D(q)  \nonumber \\
&+ \Big[2\,\vp\cdot\vq \, f_\kappa^\lambda(Q)\, \ell_\kappa(\Omega,Q) \,\ell_\kappa(\omega,q)\,k_\kappa(\omega,q) +(\vp\,^2 + \vp\cdot\vq)\,f_\kappa^\lambda(q)\, k_\kappa(\Omega,Q)(\omega^2 \,f_\kappa^\lambda(q)^2-\ell_\kappa(\omega,q)^2 ) \Big]\, \vq\,^2\, \p_\kappa S_\kappa^\nu(q)  \Bigg\} ,\label{dsfn} \\
 \p_\kappa f_\kappa^\lambda(\varpi,p) & =\displaystyle 2\frac{g_\kappa}{\varpi}  f_\kappa^\lambda(p)\int_{\omega,\vq}  \frac{\vq\,^2 +(\vp\cdot\vq)}{P_\kappa(\omega,q)^2 P_\kappa(\Omega,Q)}\Bigg\{-\Omega\,\vp\cdot\vq\, f_\kappa^\lambda(Q)^2\,P_\kappa(\omega,q)\, \p_\kappa S_\kappa^D(q)  \nonumber \\
& + 2  \Big[\Omega\,\vp\cdot\vq \, f_\kappa^\lambda(Q)^2\,k_\kappa(\omega,q)+\omega\,(\vp\,^2 + \vp\cdot\vq)\, f_\kappa^\lambda(q)^2 \,k_\kappa(\Omega,Q)  \Big]\,\vq\,^2 \, \ell_\kappa(\omega,q)\, \p_\kappa S_\kappa^\nu(q)\Bigg\} \, ,
\end{align}
\label{eqf}
\end{subequations}
\end{widetext}
with $Q\equiv |\vp+\vq|$, $\Omega\equiv \omega+\varpi$, and
\begin{subequations}
\begin{align}
\ell_\kappa(\omega, q) &= q^2 ( \fn\left(\omega, q\right)+  \nu_\kappa\, r( q\,^2/\kappa^2) ), \\
k_\kappa(\omega,q)&=\fd(\omega,q)+D_\kappa r(q^2/\kappa^2) \\
P_\kappa(\omega, q)& = \omega^2 \,\fl \left(q\right)^2 +  \ell_\kappa \left(\omega,q\right)^2 \\
S_\kappa^X (q)  &=  X_\kappa r(y)  \, \, ,\, \, y = q^2/\kappa^2 \, \, ,\, \,  X \in \{ D,\nu \}  ,\\
\kappa \partial_\kappa S^X_\kappa (y)  &= - X_\kappa \, (\eta^X_\kappa r(y) + 2 y \,\partial_{y}  r(y)),
\end{align}
\end{subequations}
and where the anomalous dimensions $\eta^X_\kappa$ are defined below.

 The breaking of the Galilean symmetry is encompassed by
  the flow of the non-linear coupling, which   can be 
  defined from the 3-point vertex function $\Gamma_\kappa^{(2,1)}$ given in \eq{gamma21} as 
\begin{equation}\label{eq:fldef}
 \lambda_\kappa= \lim_{p\to 0} \frac{4}{p^2} \Gamma_{\kappa}^{(2,1)}\left(0,\frac{\vec{p}}{2},0,\frac{\vec{p}}{2}\right)\, .
\end{equation}
The computation of the flow of $\lambda_\kappa$ is reported in \aref{app:a1}. We obtain within the NLO$_\omega$ approximation
\begin{align}
\p_s& \lambda_\kappa=- S_d\frac{2 g_\kappa}{d}\int_0^\infty \dq \int_{-\infty}^\infty \dom\frac{q^{d+3}  f_{\kappa }^{\lambda }(q)^2 }{P_{\kappa }(\omega,q)^4}\Bigg\{\nonumber\\
&\p_s S_{\kappa }^D(q) P_{\kappa}(\omega,q) \left[P_{\kappa }(\omega,q)-4 \omega ^2 f_{\kappa }^\lambda(q)^2\right]  \nonumber\\
& -4 q^2 \p_s S_{\kappa }^{\nu }(q) k_{\kappa }(\omega,q) \ell_\kappa(\omega,q) \left[P_{\kappa }(\omega,q)-6 \omega ^2 f_{\kappa }^\lambda(q)^2\right]\Bigg\}\label{eq:kdkflnlop}
\end{align}
where we  used $\int \text{d}\vq \,(\vp\cdot\vq)^2 F( q) = \frac{S_d}{d} p\,^2 \int_0^\infty \text{d}q \, q^2 F(q)$, with $S_d=2\pi^{d/2}/\Gamma(d/2)$  the $d$-dimensional solid angle. 

The flow equations within the NLO approximation can be deduced from the ones at NLO$_\omega$,
by further neglecting the frequency dependence of $f_\kappa^D$ and $f_\kappa^\nu$ in the integrands \eq{eqf} and \eq{eq:kdkflnlop}, \ie $f_\kappa^{\nu,D}(\omega,q)\to f_\kappa^{\nu,D}(q)$. 
With this replacement, the integration over the internal frequency $\omega$
 can be performed analytically (see  \cite{Kloss12} for the explicit expressions). Moreover,
 let us emphasize that one obtains in this case, after integration on $\omega$, that $\p_s \lambda_\kappa$ is exactly zero.
 This explains why the NLO$_\omega$ extension is necessary to account for a "smooth", {\it i.e } analytical, breaking of Galilean
  symmetry, as the one occurring in the SR case with the correlator $D_\tau$.
 
\begin{figure*}[ht]
\centering
\includegraphics[scale=0.8]{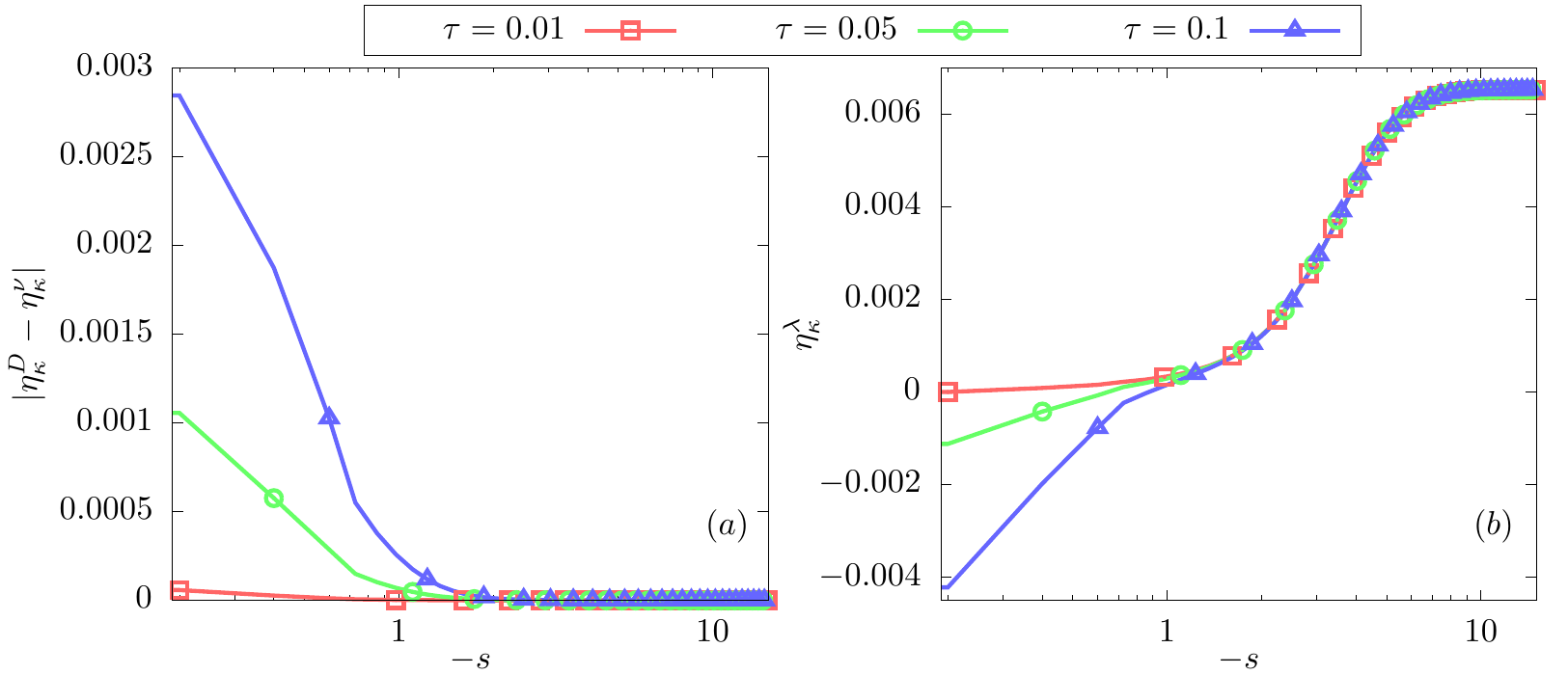}
\caption{Evolution with the RG time $s= -\log(\kappa/\Lambda)$ of (a) $\eta^\lambda_{\kappa}$ and (b) $|\eta^D_{\kappa}-\eta^\nu_{\kappa}|$, for different values of $\tau=0,0.01,0.05,0.1$. One observes that: (a) the time-reversal symmetry is dynamically restored along the flow since $\eta^D_*= \eta^\nu_*$ at the fixed point, which implies that $\chi=1/2$, (b) the Galilean invariance is almost restored: $\eta^\lambda_*$ takes a very small value for all $\tau$. As explained in the text, this residual non-zero violation of Galilean invariance is induced by the NLO$_\omega$ ansatz, which implies that   $z=2-\chi-\eta^\lambda_*$ slightly deviates (by less than 0.5\%) from the SR-KPZ value   $z=3/2$.}
\label{fig:etaSRd1}
\end{figure*}

\subsubsection{Anomalous dimensions and dimensionless flows}

The global scaling of the renormalization functions can be determined at a specific normalization point $(\varpi_{\rm NP},p_{\rm NP})$.
 This is equivalent to the choice of a prescription point in standard perturbative RG. Within the context of NPRG, this normalization point
 can be in general  simply chosen as $(0,0)$,  because the flow is regularized and no singularity occurs at vanishing momentum
 and frequency. Here, this is more subtle in the case of power-law correlations which may introduce a non-analyticity at zero frequency, since the flow is not regularized in the frequency sector.  It is useful in this case to consider a non-zero normalization frequency.
Hence, we define two scale-dependent coefficients $\nu_\kappa$ and $D_\kappa$ as the normalizations of $f_\kappa^\nu$ and $f_\kappa^D$ at the 
point $(\varpi_0,0)$ according to
\begin{equation}
\label{eq:defexpo}
 D_\kappa \equiv \fd(\varpi_0,0)\; ,\quad \quad \nu_\kappa \equiv \fn(\varpi_0,0)\, ,
\end{equation}
where  $\varpi_0=0$ unless stated otherwise.
We emphasize that the choice of the normalization point is in principle arbitrary and the results should not depend on it.
 This is true in the exact theory. However, as for the choice of the cutoff function,  once
 approximations are performed, one can expect a small residual dependence on the precise value of the normalization point. We
 checked that it is negligible, see  \aref{APP-power}.

The two coefficients $D_\kappa$ and $\nu_\kappa$ encompass the renormalization of the fields and the  scaling between space and time.
Their flow can be simply obtained from the limit $(\varpi,p)\to(\varpi_0,0)$ in \Eq{dsfd} and \Eq{dsfn} respectively.
One can define two running scaling dimensions associated with these coefficients as
\begin{equation}
 \eta^D_\kappa = -\kappa \p_\kappa \ln D_\kappa \; ,\quad \quad\eta_\kappa^\nu = -\kappa \p_\kappa \ln \nu_\kappa\, .
\end{equation}
One can show that the critical exponents can be expressed in terms of the fixed point values of these scaling exponents as
\cite{Canet11a}
\begin{equation}
 z = 2-\eta_*^\nu \; ,\quad \quad \chi = (2-d+\eta_*^D-\eta_*^\nu)/2\, .
\label{eq:expo}
\end{equation}
For $f_\kappa^\lambda$, the shift-gauged symmetry imposes that $\fl(\varpi,0)=1$ for all $\varpi$, and in particular for $\varpi_0$, so
 this function does not introduce another scaling coefficient.

In the following, we are interested in the fixed points of the RG flow equations. Indeed, a fixed point means that all quantities
  do not depend on the scale $\kappa$ any longer, which physically implies that the system is scale invariant,  critical. As common in RG approaches, the 
 appropriate way to search for a fixed point is to switch to dimensionless quantities.
We hence define dimensionless momenta, \eg $\hat p=p/\kappa$, and frequencies, \eg $\hat \varpi=\varpi/(\nu_\kappa\kappa^2)$, and  consider  the dimensionless functions obtained as $\hat f_\kappa^X(\hat \varpi,\hat p) = f_\kappa^X(\varpi,p)/X_\kappa$. Their flow equation is thus given by
\begin{align}
 \p_s \hat f_\kappa^X(\hat \varpi,\hat p)&= \big[ \eta_\kappa^X +(2-\eta_\kappa^\nu) \hat \varpi \p_{\hat\varpi}+\hat p \p_{\hat p} \big]\hat f_\kappa^X(\hat \varpi,\hat p)\nonumber\\
 & + \hat I_\kappa^X (\hat \varpi,\hat p) \label{adim}
\end{align}
where $\hat I_\kappa^X$ is the non-linear part of the flow equations $\p_s f_\kappa^X/X_\kappa$, given in \eq{eqf}, 
expressed in terms of dimensionless variables, and with $X_\kappa=D_\kappa,\nu_\kappa,1$ and
 $\eta_\kappa^X=\eta_\kappa^\nu,\eta_\kappa^D,0$ for $f_\kappa^D$, $f_\kappa^\nu$ and $f_\kappa^\lambda$ respectively.

 Finally, we denote the flow of $\lambda_\kappa$ as
\begin{equation}
  \kappa\p_\kappa \ln \lambda_\kappa = -\eta_\kappa^\lambda .
\end{equation}
The flow of the dimensionless coupling $\hat g_\kappa \equiv \kappa^{d-2}\lambda_\kappa^2 D_\kappa/\nu_\kappa^3$ can be expressed as 
\begin{equation}
\p_s \hat{g}_k =\hat{g}_k\left( d-2-2\eta_\kappa^\lambda+3\eta_\kappa^\nu-\eta^{D}_\kappa \right) \, .
\label{dsg}
\end{equation}
At a non-gaussian fixed point $\hat{g}_k\neq 0$, this implies the exact relation
\begin{equation}
 z+\chi - 2 =  \eta_*^\lambda \, .
\end{equation}
Hence, if Galilean symmetry is present, then  $\eta_*^\lambda=0$ and one recovers the standard relation $z+\chi=2$.
A non-zero $\eta_*^\lambda$ quantifies the violation of Galilean invariance.

\subsubsection*{Flow of $w_\kappa^\theta$ in the NLO approximation}

The presence of a power-law noise correlator $D_\infty(\omega,\vq)$ in \eq{eq:choiseD}  introduces another coupling $w_\kappa^\theta$   related to the non-analytic part.
 The function $\fd$ is now composed of two parts
\begin{equation}
 \fd(\omega,q) = \tilde{f}^D_\kappa(\omega,q) + w_\kappa^\theta \omega^{-2\theta}.
\end{equation}
In principle, the NPRG flow is analytic, such that no non-analytic contribution can arise to renormalize the coupling $w_\kappa^\theta$.
 The situation is more subtle here since the frequency sector is not regularized, see discussion in \aref{APP-power}, but 
  the  non-renormalization of $w_\kappa^\theta$ is preserved.
 Defining the dimensionless running coupling $\hat{w}_\kappa^\theta$ as
\begin{equation}
\hat{w}^\theta_\kappa=\kappa^{-4\theta} w_\kappa^\theta \frac{1}{D_\kappa\nu_\kappa^{2\theta}}\, 
\end{equation}
one obtains its flow as
\begin{equation}
\p_s \hat{w}^\theta_\kappa = \hat{w}^\theta_\kappa \left( -4\theta+\eta_\kappa^{D} +2 \theta \eta_\kappa^\nu \right)\, .
\label{dsw}
\end{equation}
For any fixed-point solution for which $ \hat{w}^\theta_*\neq 0$, one deduces that
\begin{equation}
 \eta_*^{D} = 4\theta-2 \theta \eta_*^\nu
\end{equation}
which yields if $\hat g_*\neq0$
\begin{equation}
 \eta_*^\lambda = \frac{1}{2} (2-d+4\theta -(3+2\theta)\eta_*^\nu),
\end{equation}
which is non-zero in general. Hence, if a LR fixed-point with $\hat{w}^\theta_*\neq 0$ exists and is stable,
 it is associated with a violation of Galilean symmetry. Assuming that the two fixed-points, the  LR and the SR ones, exist and compete,
  then the transition from one to the other occurs when the corresponding dynamical exponents are equal, that is
  for $z_{\rm LR}= z_{\rm SR}$ and thus $\eta_\lambda^*=0$. One deduces  that the corresponding
     critical value ${\theta_{\textrm{th}}}$ is given by
\begin{equation}\label{eq:thetac}
{\theta_{\textrm{th}}}(d)=\frac{1}{2(\eta_\nu^*-2)}(2-d-3\eta_\nu^*)\, .
\end{equation}
One then expects a transition from a SR to a LR
 dominated phase with critical exponents satisfying:
\[\begin{matrix}
\hbox{SR}: & z+\chi = 2, & \theta<{\theta_{\textrm{th}}} \\
\hbox{LR}: & z+\chi = 2 -\eta_\lambda^*(\theta), & \theta>{\theta_{\textrm{th}}} \,{.}
\end{matrix}\]

\section{Results}
\label{sec:RES}

In this section, we only consider dimensionless quantities, so we omit the hat symbols to alleviate notations.

\begin{figure*}[ht]
\centering
\includegraphics[scale=0.8]{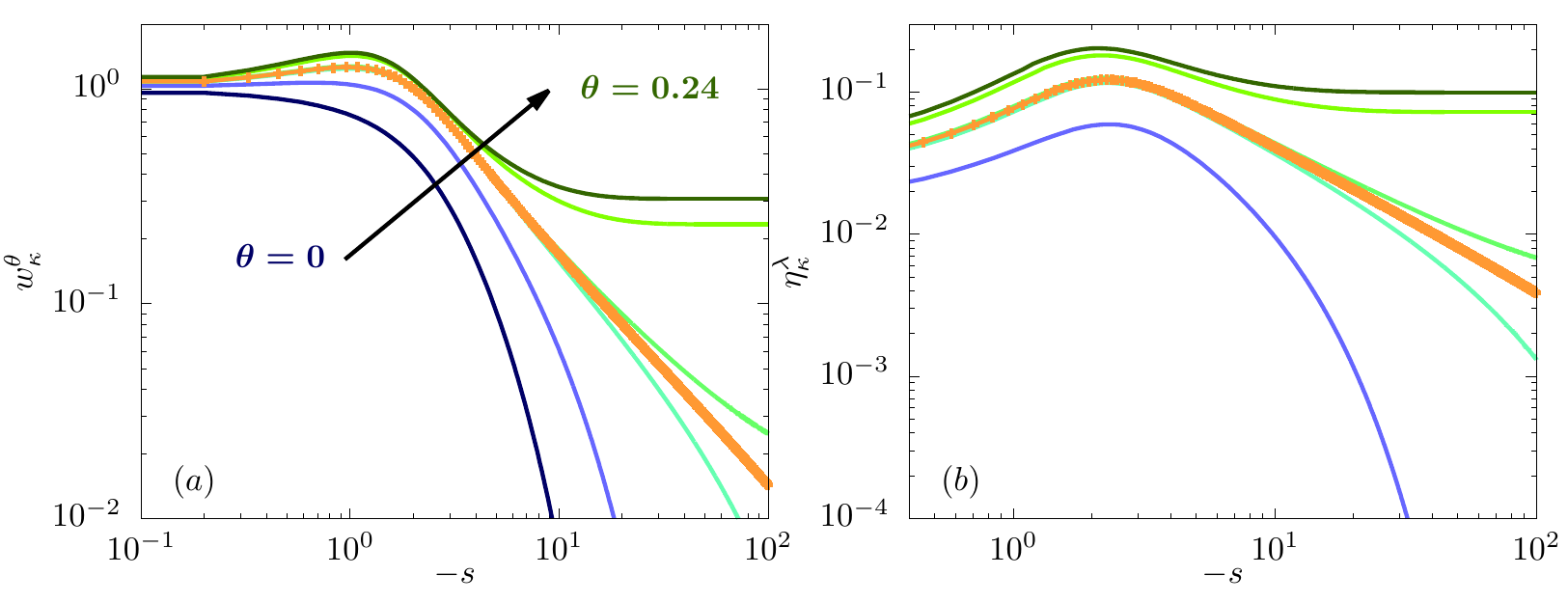}
\caption{Evolution with the RG time $s$ in $d=1$ of (a) the LR coupling $w^\theta_\kappa$ and (b) the violation of Galilean invariance $\eta^\lambda_{\kappa}$, for different values of $\theta=0,0.1,0.16,0.166,0.17,0.22,0.24$ (from bottom to top) in $d=1$. For $\theta<{\theta_{\textrm{th}}}$, the flow reaches the SR-KPZ fixed point with $w^\theta_*=0,\eta^\lambda_*=0$, while for $\theta>{\theta_{\textrm{th}}}$,  a LR  fixed point with $w^\theta_*\neq 0,\eta^\lambda_* \neq 0$ is reached. The critical value is ${\theta_{\textrm{th}}}=0.166$ confirming the theoretical prediction ${\theta_{\textrm{th}}}=1/6$. Interestingly, it is clearly identified as  the value leading to an algebraic decay of $w^\theta_\kappa$ and  $\eta^\lambda_{\kappa}$ in the RG time $s$ (bold orange line).}
\label{fig:dthetad1}
\end{figure*}

\subsection{Temporal correlations with a finite correlation time}
\label{sec:sr}

We consider the KPZ action with the microscopic noise correlator $D_\tau$ defined in \eq{eq:choiseD}.
As explained previously, $\Gamma_\kappa$ coincides with the microscopic action at the microscopic scale $\kappa=\Lambda$.
Comparing \eq{eq:KPZaction} and \eq{eq:nlopansatz}, one concludes that  this initial condition corresponds to
\begin{equation}
f^D_{\kappa=\Lambda}(\varpi,p)=D_\tau(\varpi,p)= {e}^{-\frac{1}{2}\varpi^2\tau^2}
\label{initfd}
\end{equation}
and 
\begin{equation}
 f_\Lambda^\nu(\varpi,p)= 1,\quad \quad f^\lambda_\Lambda(\varpi,p)\equiv 1\, .
 \label{eq:CI}
\end{equation}
Hence at the microscopic level, both the Galilean invariance (since $f^D_{\Lambda}$ depends on frequency) and the time-reversal symmetry in $d=1$ 
 (since $f^D_{\Lambda}\neq f_\Lambda^\nu$) are broken.

 Let us focus on $d=1$. In this dimension, the function $f^\lambda_\kappa$ is
  kept to one as imposed by the time-reversal symmetry. 
 We integrated numerically the flow equations for the two functions $f^D_\kappa(\varpi,p)$ and $f^\nu_\kappa(\varpi,p)$ (NLO$_\omega$ approximation), together with the flow equations for the coupling $g_\kappa$, and for the coefficients
 $\nu_\kappa$ and $D_\kappa$,
 for different initial values of $\tau$ between $0$ and $1$. Details on the numerical procedure are provided in \aref{sec:num}.

 For all values of $\tau$, we observed that the flow reaches a fixed point, with stationarity in $\kappa$ for all quantities. The coupling $g_\kappa$ tends to a fixed-point value $g_*$. At the same time, the renormalization functions $f^D_\kappa$ and $f^\nu_\kappa$ smoothly evolve to endow a fixed-point form, which does not depend on the value of $\tau$, as illustrated for $f^D_\kappa$ in \fref{fig:fomSRd1} (a). This means that  the large distance physics is universal, \ie independent of the microscopic details,  and  it corresponds to the SR-KPZ universality class (the same fixed-point is attained as for $\tau=0$).

 Furthermore, although they start with very different shapes, the two functions $f^D_\kappa$ and $f^\nu_\kappa$ become equal at the fixed point
 $f^D_*(\varpi,p)\equiv f^\nu_*(\varpi,p)$, as illustrated in \fref{fig:fomSRd1} (b). This means that the time-reversal symmetry is dynamically restored at large distances. This  is further illustrated in \fref{fig:etaSRd1} (b),
  which shows that the difference $|\eta_\kappa^\nu-\eta_\kappa^D|$ vanishes at the fixed point for all $\tau$. According to \Eq{eq:expo}, this implies that the $\chi$ exponent is exactly the SR-KPZ one $\chi_{\rm SR}=1/2$.

Moreover,  the Galilean symmetry is also restored at the fixed point, although only approximately at NLO$_\omega$. This can be assessed by the value of
 $\eta^*_\lambda$, which is represented in \fref{fig:etaSRd1} (a). One observes that it reaches a constant value, which is not strictly zero but a small number of order 0.0065. As explained before, this reflects the spurious
 violation of Galilean invariance induced by the NLO$_\omega$ ansatz (dependence in $\p_t$ rather than $\tilde D_t$).
 This value is the same as for the pure SR-KPZ case (for $\tau=0$) and yields an error of less than 0.5\% on the exponent $z$. The NLO$_\omega$ approximation
 is hence still accurate despite its simplification compared to SO. Furthermore, we observe  that for any finite $\tau$, the function $f_*^D$ decays at large frequency as a power law ${f^D_{*,\tau}(\varpi,0) \sim \varpi^{-\eta^D_*/z}}$, with $z=2-\chi-\eta^\lambda_*$, very close to the pure SR-KPZ case ${f^D_{*,\tau=0}(\varpi,0) \sim \varpi^{-1/3}}$. Hence one can conclude that for all $\tau$,  the universal properties of the interface are the standard SR-KPZ ones.

In two dimensions, the NLO$_\omega$ approximation does not seem to suffice to properly describe the pure SR-KPZ case.
 We did not succeed in accessing the fixed point within this scheme, probably because the violation of Galilean symmetry 
 induced by the ansatz (through neglecting all higher-order vertex functions) is too severe in $d=2$. On the other hand, the NLO scheme alone does not allow to implement an initial condition which involves a  functional frequency dependence of $f^D_\Lambda$ as in \eq{initfd}. Hence, to study the effect of a temporal SR-correlated noise in $d=2$ would require to use the full SO ansatz (which does not induce any artificial violation of Galilean symmetry). This is beyond the scope of this work. We will thus restrict in $d=2$ to the study of the LR case, which can be studied at NLO.

 To summarize on the temporally SR-correlated noise, we found in $d=1$ that its presence 
  does not change the large-distance and long-time properties of the interface, which is still characterized by the SR-KPZ
 universality class. Hence, 
 although both the Galilean and time-reversal symmetries are broken at the microscopic level, these symmetries are restored dynamically along the flow.
  This is the first  analysis of the effect of   SR time-correlations in the KPZ equation, which is here rendered possible by the both functional and non-perturbative formalism we use.

\begin{figure}[t]
\includegraphics[scale=0.8]{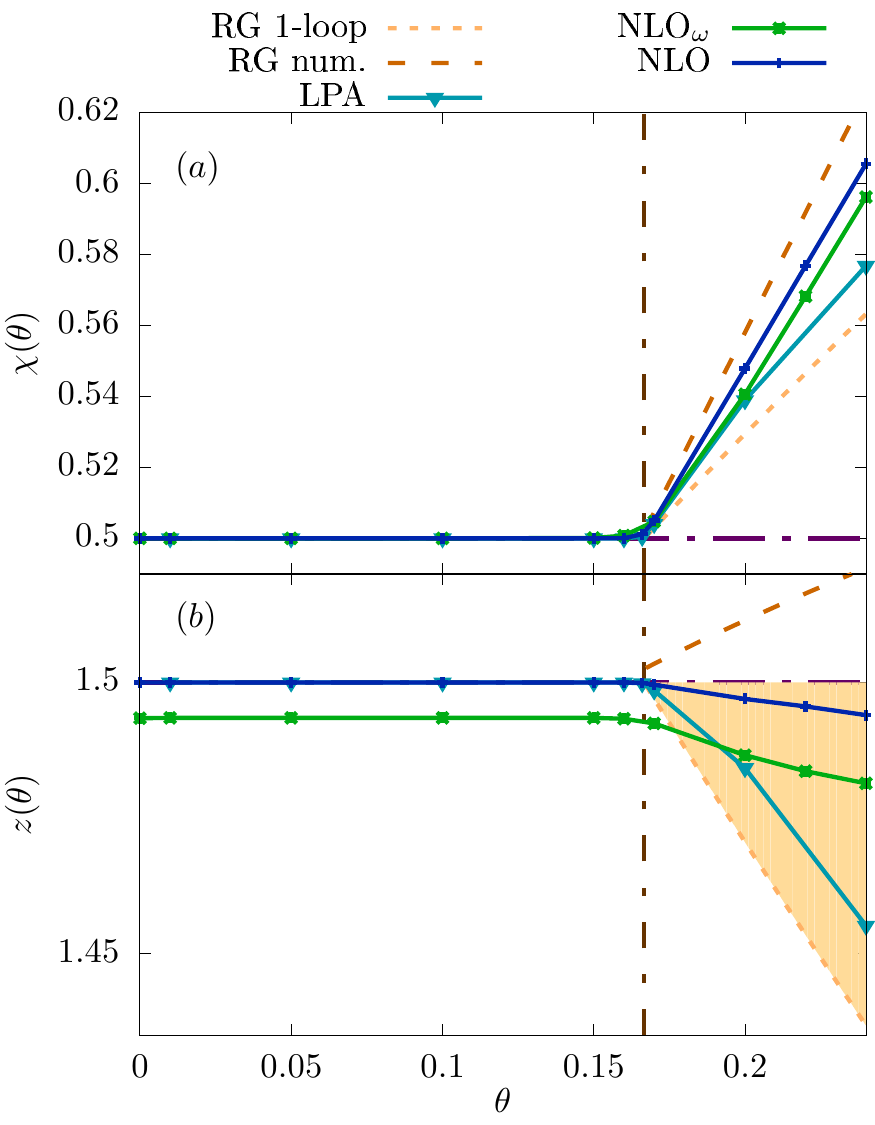}
\caption{(a) Roughness exponent $\chi$ and (b) dynamical exponent $z$ as a function of $\theta$ in $d=1$ and for different approximation schemes. The exponents take the SR-KPZ values $\chi=1/2$ and $z=3/2$ up to a critical value of $\theta$ very close to the theoretical prediction ${\theta_{\textrm{th}}}=1/6$. Beyond this value, the exponents vary continuously with $\theta$. The LPA results  improves the analytical RG result at one-loop given in \eqref{exp-oneloop}, where we recall the authors set $\eta^\lambda_\kappa=0$, but still differ from the NLO results. The NLO and NLO$_\omega$ results are very close, despite the slight shift in $z$ at NLO$_\omega$ due the the residual breaking a Galilean invariance within this sheme. At NLO, we find an almost linear behavior of $\chi(\theta)\simeq 1.43\theta+0.26$.
 It turns out to be reasonably  close to numerical estimation from the approximate RG equations \eqref{exp-drg}  for $\chi$, but not for $z$.  All the estimates from NPRG lie within the bounds given in \eqref{eq:fedbounds} (shaded region).} 
\label{fig:expd1}
\end{figure}

\begin{figure*}[ht]
\centering
\includegraphics[scale=0.8]{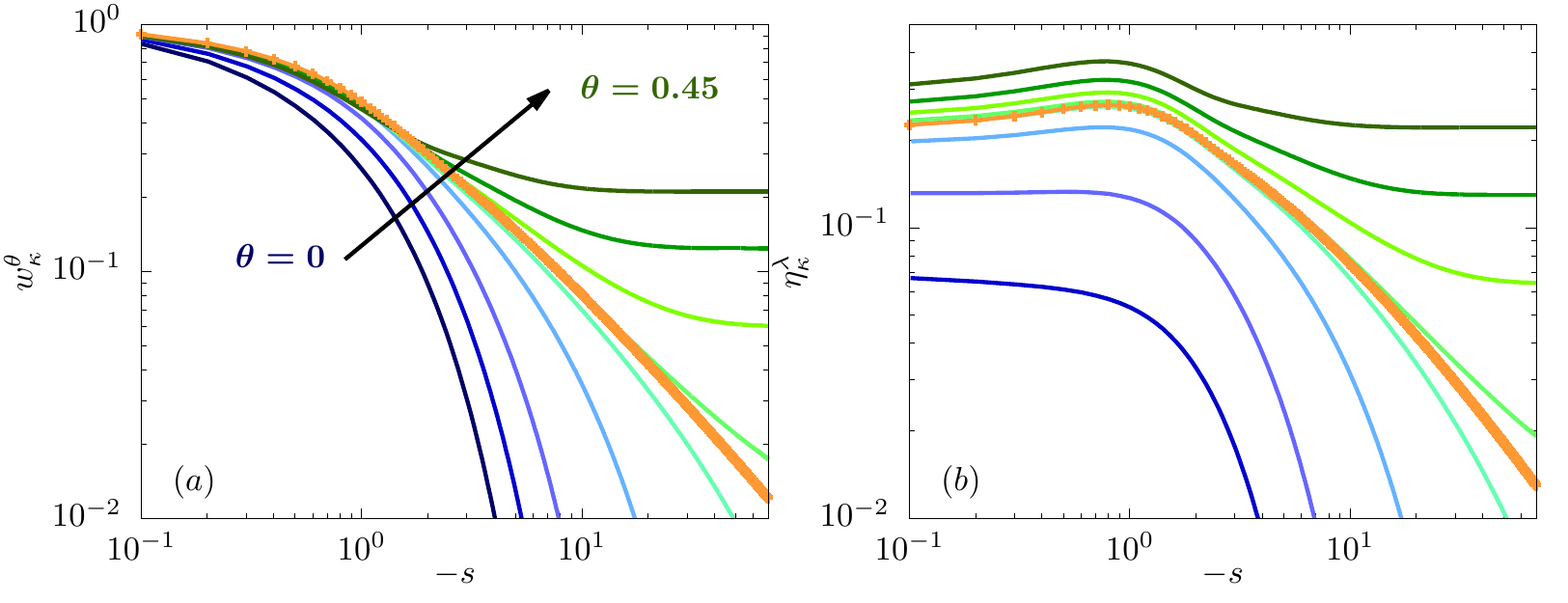}
\caption{(a) The LR coupling $w^\theta_\kappa$ and (b) $\eta^\lambda_{\kappa}$, for different values of $\theta=0,0.1,0.2,0.3,0.34,0.35,0.4,0.45$ (from bottom to top) and  ${\theta_{\textrm{th}}}=0.346$ (bold orange line) as a function of the   RG time $s$, in $d=2$. Two distinct behaviors, corresponding to the SR and the LR fixed points are observed, 
 separated by the critical value ${\theta_{\textrm{th}}}$ for which  $w^\theta_\kappa$ and $\eta^\lambda_\kappa$ vanish algebraically with  $s$.}
\label{fig:dthetad2}
\end{figure*}

\begin{figure}[ht]
\includegraphics[scale=0.8]{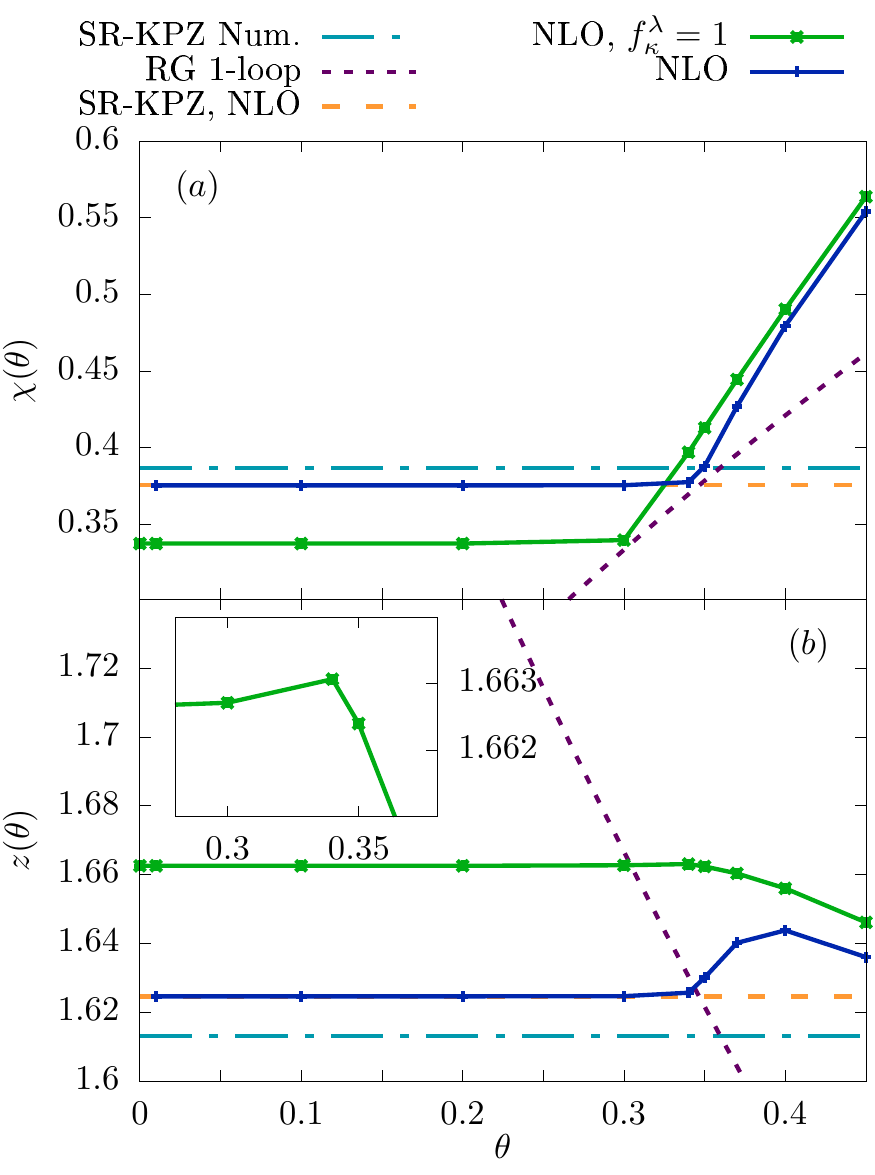}
\caption{(a) The roughness $\chi(\theta)$ and (b) dynamical $z(\theta)$ critical exponents as a function of the LR exponent $\theta$ in $d=2$, and for different approximation schemes. 
 The reference values for the pure KPZ case, obtained  within NLO and from recent numerical simulations \cite{Pagnani15}, are represented as the dashed lines SR-KPZ NLO and SR-KPZ Num., respectively. 
 We find in $d=2$ at NLO two regimes, as in $d=1$. The critical exponents coincide with the SR-KPZ ones below a critical value ${\theta_{\textrm{th}}} \simeq 0.346$, while beyond this value, we obtain $\theta$-dependent critical exponents.
  Within the simplified NLO approximation with $f_\kappa^\lambda=1$, the qualitative picture is the same, although the curves are shifted because the values for the exponents at the SR-KPZ fixed point  differ a bit (less than 10 \%) within this scheme.
The one-loop analytical result \eq{exp-oneloop} 
 from DRG is also represented for comparison, although in this case,  the value for ${\theta_{\textrm{th}}}$ cannot be obtained from the perturbative analysis.  It can be estimated here as the intersection between this prediction and the SR-KPZ values for the exponents.} 
\label{fig:expd2}
\end{figure}

\subsection{Power-law temporal correlations}

We now investigate the presence of correlations in the microscopic noise with 
  no typical length-scale, $i.e.$ the power-law LR correlations $D_\infty$ in \eq{eq:choiseD}. 
 This corresponds to the initial condition
\begin{equation}
f^D_{\kappa=\Lambda}(\varpi,p)=D_\infty(\varpi,p)= 1 + w^\theta_\Lambda \varpi^{-2\theta}
\end{equation}
 together with \eq{eq:CI}. As explained in  \sref{sec:flow}, this LR noise introduces the new dimensionless coupling constant $w_\kappa^\theta$ and two different scenarii may now emerge: either the SR part of the noise dominates,
  corresponding to a stable SR fixed point with $w_*^\theta=0$, or the LR part dominates, 
  corresponding to a stable LR fixed point with   $w_*^\theta\neq0$. Since for such a fixed point Galilean
   symmetry is broken,  $z+\chi\neq 2$,  there is no simple way to compute the associated LR critical exponents even in $d=1$.

To study the LR noise, it is enough to work within the NLO approximation, since the analytical dependence in frequency of the  functions $f_\kappa^X$ is not essential (sub-dominant) in this case.
 The advantage is that there is no spurious (\ie introduced by the ansatz) breaking of Galilean symmetry at NLO. Indeed, the analytical part of the flow of $\lambda_\kappa$ vanishes at NLO as explained in \sref{sec:flow}.
 The only contribution to the flow of $\lambda_\kappa$ thus stems from the non-analytical part of $f^D_\kappa$ and reads
\begin{align}
\p_s \lambda_\kappa&=S_d\frac{8 g_\kappa w^\theta_\kappa}{d}\int_0^\infty \dq q^{d+5}  f_{\kappa }^{\lambda }(q)^2\p_s S_{\kappa}^{\nu}(q)\ell_\kappa(q) \nonumber\\
&\times\int_{-\infty}^\infty \dom \frac{\omega^{-2\theta}}{P_{\kappa }(\omega,q)^4} \left[P_{\kappa }(\omega,q)-6 \omega^2 f_{\kappa}^\lambda(q)^2\right]\label{eq:etalnlolr} \, .
\end{align}

In this part, we also use the local potential approximation, in order to compare the results stemming from successive orders of approximations.
The LPA flow equations can be simply deduced from the NLO ones by completely neglecting the momentum
 and frequency dependence of the running functions, which is thus equivalent to simply considering the flow of the two dimensionless couplings $g_\kappa$ and $w_\kappa^\theta$, 
 and the two anomalous dimensions $\eta_\kappa^D$ and $\eta_\kappa^\nu$.

\subsubsection{One dimensional case}

As for the SR correlated noise, we fix  $f^\lambda_\kappa=1$ in $d=1$ in order to satisfy the time-reversal symmetry. 
We integrated numerically the NLO flow equations for $f_\kappa^D$ and $f_\kappa^\nu$ together with the flow equations for 
the two dimensionless couplings $g_\kappa$ and $w_\kappa^\theta$ and for the two anomalous dimensions. We find two distinct regimes depending on the 
value of $\theta$, as illustrated on \fref{fig:dthetad1}. For $\theta<{\theta_{\textrm{th}}} = 1/6$,  $ g_\kappa$ flows 
to a finite fixed point value $g_*$ while $w^\theta_\kappa$ flows to zero. At the same time, 
$\eta^\lambda_\kappa$ also flows to zero, hence Galilean invariance is dynamically restored,
and the critical exponents take the SR-KPZ values $\chi_{\rm SR}=1/2$ and $z_{\rm SR}=3/2$.
 The long-distance physics is hence the same for all $\theta<\theta_{\rm th}$ and controlled by the SR-KPZ fixed-point.
  For  $\theta>{\theta_{\textrm{th}}}$, both  $g_\kappa$ and  $w^\theta_\kappa$ flow to a non-zero fixed-point value, and the violation of Galilean invariance $\eta^\lambda_*$ increases with $\theta$, as illustrated on \fref{fig:dthetad1}. Hence in this regime, the long-distance properties are controlled by a line of LR fixed points, with critical exponents depending on $\theta$. The critical value ${\theta_{\textrm{th}}}=1/6$ delimiting the two regimes is clearly identified on \fref{fig:dthetad1} by the algrebraic decay of $w_\kappa^\theta$ and $\eta^\lambda_\kappa$ with the RG time $s$. 

These findings are in agreement with the results presented in \cite{Medina89,Fedorenko08}, and show that a pure SR-KPZ regime is not destroyed for an infinitesimal $\theta$, contrary to the scenario advocated by SCE or Flory approaches. 
The critical exponents $\chi$ and $z$ obtained at NLO are represented on \fref{fig:expd1},
 and compared to the DRG approach of \cite{Medina89} for which explicit results are given.

We also performed the same analysis within different approximations of NPRG to test the robustness of the results.
 Within the NLO$_\omega$ scheme, the existence of the two regimes is confirmed. However,
  since in this approximation a residual breaking of Galilean invariance even at $\theta=0$  ($\eta_*^\lambda \simeq 0.0065$) induced
  by the ansatz  subsists at the SR fixed point, the critical value of ${\theta_{\textrm{th}}}$ is slightly shifted,
   but the critical exponents are close to the NLO ones, see \fref{fig:expd1} (especially if
    the small shift 0.0065 is compensated for). 

 Within the LPA, we also find the two regimes with the same critical value ${\theta_{\textrm{th}}}$, but the critical exponents slightly differ, they lie closer to the one-loop results \eqref{exp-oneloop} as could be expected. The NLO results fall in between the numerical approximation of \cite{Medina89}
 and the LPA results. Let us emphasize that all our estimates for $z_{\rm LR}(\theta)$  are decreasing with $\theta$ and  lie within the bounds \eq{eq:fedbounds} derived in \cite{Fedorenko08}. 

Let us finally note that these theoretical results are not in agreement with the most recent numerical simulations \cite{ales2019}.
 In the simulations, the SR-KPZ phase is destroyed for any non-zero $\theta$ (no threshold value) and the value of the dynamical critical
 exponent is found to be constant $z\simeq 3/2$ independently of $\theta$. We have no clear understanding of these discrepancies. 
  The main differences lie in the fact that the simulations deal with finite-size systems in discretized space and time whereas we work in the infinite-size limit 
in continuous space-time. In general, finite-size effects tend to smear out transitions. A putative explanation could be that very large system sizes are required to clearly resolve  the  threshold. As for the value of $z$, we emphasize that the deviation from the SR value $z_{\rm SR}=3/2$  in the LR phase is small, as manifest in \fref{fig:expd1}(b) [more quantitatively,  $z_{\rm LR}=1.42(5)$ for $θ=0.47$ and $z_{\rm LR}=1.49(4)$ for $θ=0.24$]. Indeed, $z$ only depends on $\eta_*^\nu$, which is less sensitive than $\eta_*^D$ to the presence of the LR noise. Since $z$ is determined in \cite{ales2019} from the collapse of the structure function, it is possible that the numerical results are in fact still compatible with such small variations. These points deserve further studies.

\subsubsection{Two dimensional case}

\label{sec:d2}

In two dimensions, the estimation of $\chi$ for the pure SR-KPZ within the NLO approximation is $\chi_{\rm SR}\simeq 0.375$, in good agreement with the value $\chi\simeq 0.3869(4)$ from  recent numerical simulations \cite{Pagnani15}.
 Substituting the NLO value of $\eta^\nu_*\equiv 2-z=\chi$ in  \eqref{eq:thetac}, one obtains a theoretical estimate of the critical  value ${\theta_{\textrm{th}}}=0.346$. 
In the work of \cite{Medina89}, a non-physical divergence in $\omega$ appears for $\theta\geq 1/4$. Within the present work, {a similar singularity at $\theta=1/4$ arises in the limit of zero frequency.}
 It occurs in the flow equation of $f^D_\kappa$, in the term proportional to $\p_\kappa S^D_\kappa$,
 and is  proportional to $(\omega+\varpi)^{-4\theta}$. The singularity is thus present only  at zero external frequency $\varpi$,
 that is for the calculation of the anomalous dimension $\eta_\kappa^D$. 
  {This residual divergence is due to the absence of a regulator in the frequency sector.} If we could use a frequency-dependent regulator which do not break Galilean invariance, this problem would not exist. Without such a regulator, 
 this problem can be nevertheless avoided by shifting the normalization point $\varpi_0$ of the anomalous dimensions \eq{eq:defexpo}, to a non-zero external frequency  (see \aref{sec:num}). Hence, it can be quite simply dealt with, at variance with perturbative RG.

 We performed the same analysis as for the one-dimensional case. We observed that for $\theta<{\theta_{\textrm{th}}}$, the Galilean invariance is restored by the flow, and the large distance physics is described by the pure SR-KPZ fixed point. This is illustrated on \fref{fig:dthetad2} which shows that the LR coupling $w_\kappa^\theta$ vanishes at the fixed-point, and $\eta_\kappa^\lambda$ also vanishes (exactly at NLO). For $\theta>{\theta_{\textrm{th}}}$,  the LR coupling $w_\kappa^\theta$ 
 and $\eta_\kappa^\lambda$ both reach a non-zero fixed-point value, which depends on $\theta$. This corresponds to a LR fixed-point, where Galilean invariance remains broken. The critical value ${\theta_{\textrm{th}}}$ can be identified  on \fref{fig:dthetad2} by the algebraic decay of the flow, separating the two different behaviors.

The results for the critical exponents are shown on Fig. \ref{fig:expd2}, 
within two versions of NLO: either with a non-trivial flow for $f^\lambda_\kappa(p)$, or setting  $f^\lambda_\kappa(p)=1$. 
This latter scheme is similar to the one used in $d=1$, although it is imposed in this dimension by the time-reversal symmetry, whereas it is arbitrary in $d=2$. Both results are in agreement.

\subsection{Anomalous scaling at the LR fixed point}

A recent numerical study of the KPZ equation with power-law time correlation in $d=1$ by Al\'es and L\'opez \cite{ales2019} unveiled some anomalous scaling 
 at large values of $\theta$. In the NPRG framework, anomalous scaling can emerge as a consequence of a non-decoupling of the low- and high-momentum modes in the flow equations.
 This was indeed evidenced in the context of  turbulence from a study of the stochastic Navier-Stokes equation for incompressible flows \cite{Canet16,Tarpin17} and
 was related in this case to intermittency.
 
 Let us investigate this point for KPZ with LR correlations, focusing in $d=1$. For this, we
  study  in more depth the scaling properties at the fixed point. The fixed point equation is given by \eqref{adim} with $\p_s f_*^X=0$, that is 
\begin{equation}\label{eq:adimfp}
- I^X_* (\vpi,p)= \big[\eta_*^X +(2-\eta_*^\nu)  \varpi \p_{\varpi}+ p \p_{ p} \big] f_*^X( \varpi, p).
\end{equation}
For standard scale invariance,   the non-linear part of the flow $ I^X_\kappa$ becomes negligible compared with $f_*^X( \varpi, p) /p^{-\eta_*^X}$  at large frequency and/or momentum, which
 means that the large momentum and frequency sector decouples from the low one. As a consequence,  the fixed-point universal (IR) properties are insensitive to the microscopic (UV) details. One can
  straightforwardly  show
 that the solution of  the fixed-point equation \eq{eq:adimfp} when this decoupling property is satisfied is the scaling form \cite{Canet11a}
\begin{equation}\label{eq:scalfx}
 f^X_*(\vpi,p)= p^{-\eta^X_*} \zeta^X\left(\frac{\vpi}{p^z}\right)\,,\quad  \vpi, p \gg 1\,,
\end{equation}
where $\zeta^X$  has the asymptotic behavior
\begin{equation}
\zeta^X(y)=\left\{ 
\begin{matrix}
\zeta^X_0 \,, & y\rightarrow 0 \\
\zeta^X_\infty y^{-\eta^X_*/z}\,, & y \rightarrow \infty
\end{matrix}
\right. \,.
\end{equation}
One can then show that the physical (dimensionful) two-point correlation function, which 
can be expressed in terms of the $f_*^X( \varpi, p)$,  also endows a scaling form \cite{Canet11a,Kloss12}.
Thus, for standard scale invariance, the scaling of the dimensionless functions $ f^X_\kappa$ in the large-frequency and
 large-momentum sector is fixed by the scaling dimensions $\eta^X_*$,
 calculated from the behavior in $\kappa$ of the coefficient $D_\kappa$ and $\nu_\kappa$  
 (defined in the opposite sector  $(\varpi,p)=(\varpi_0\simeq 0,0)$).
 
\begin{figure*}[ht]
\includegraphics[scale=1]{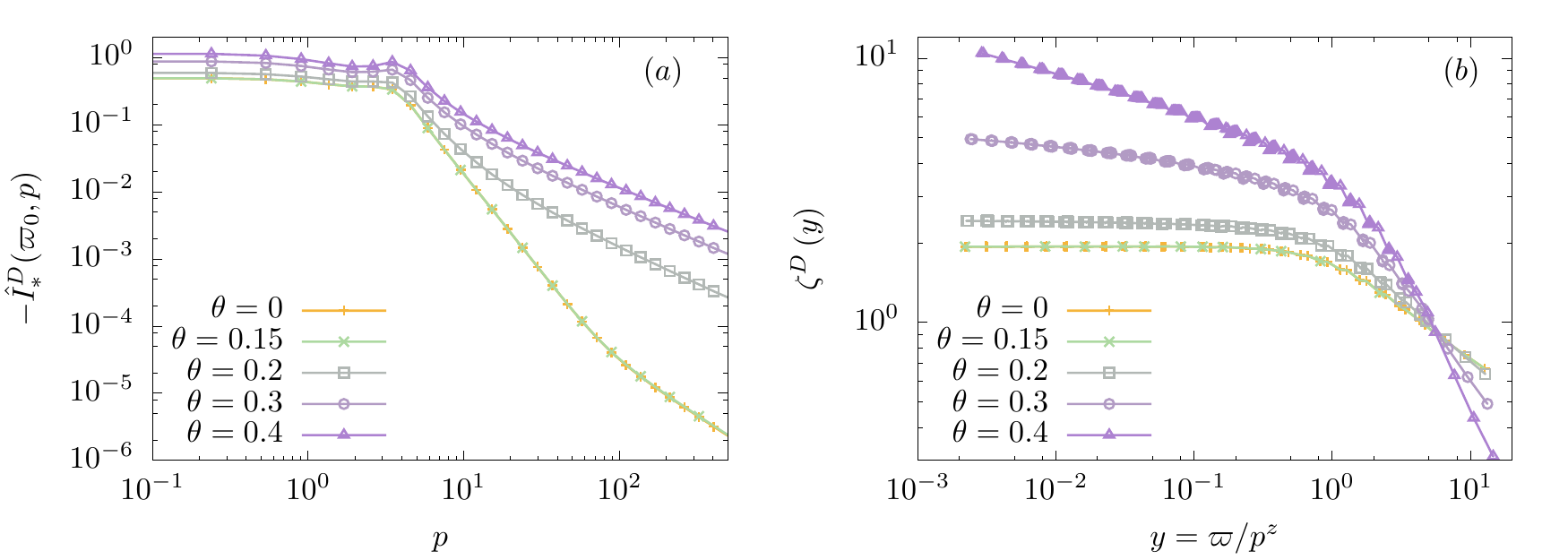}
\caption{(a) Non-linear part $I_*^D(\varpi_0,p)$ of the flow of $ f^D_\kappa$  at the fixed point as a function
 of the momentum $p$. For $\theta=0.4$, the contribution of $I_*^D$ in the $p\gg 1$ regime is almost three orders 
 of magnitude larger than in the SR-KPZ case $\theta<\theta_{\text{th}}$, which indicates that the decoupling is much 
 less effective. (b) Scaling function $\zeta^D$ associated with  $f^D_\kappa$ as a function of the scaling 
 variable $y=\varpi/p^z$. The enhanced contribution of the non-linear part of the flow at large $\theta$  translates
  into a non-zero slope in the $y\rightarrow 0$ limit. The curves for $\theta=0$ and $\theta=0.15$, which both
   belong to the SR-KPZ phase, are superimposed.} 
\label{fig:intermscal}
\end{figure*}

\begin{figure*}[ht]
\includegraphics[scale=1]{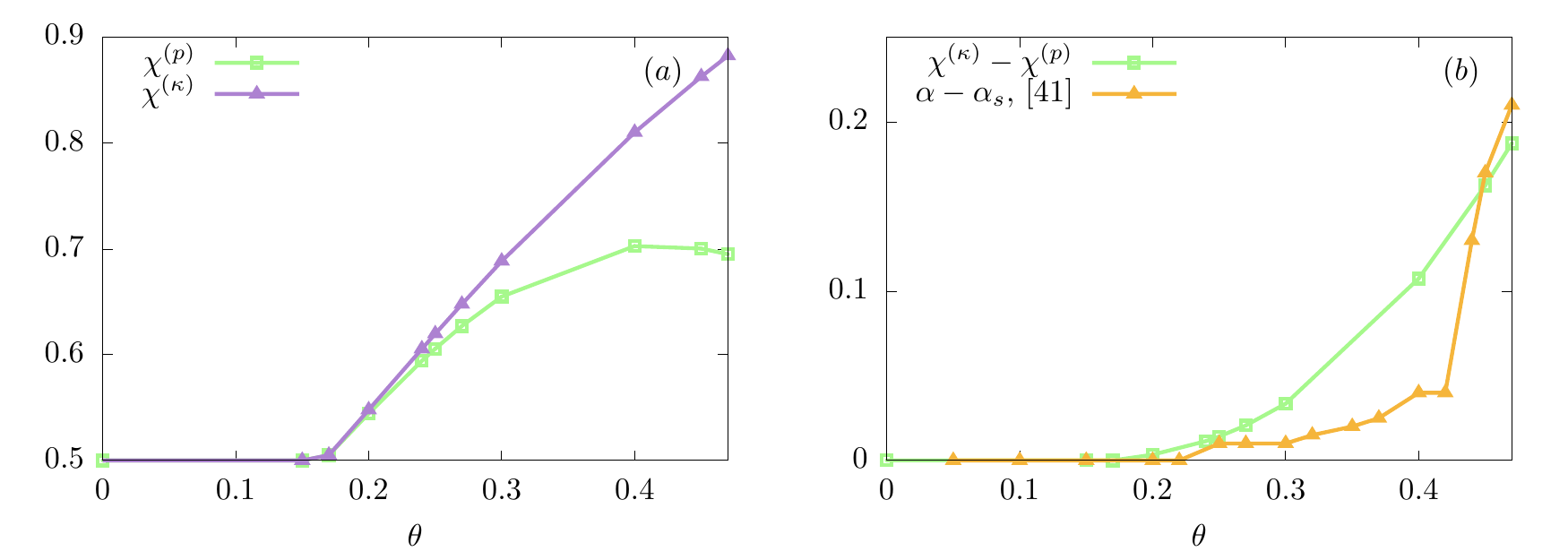}
\caption{(a) Critical roughness exponent: $\chi^{(\kappa)}$ determined from the scaling dimension of the 
renormalization functions $f_\kappa^X$ and  $\chi^{(p)}$ determined from their large momentum behavior.
 We observe that while $\chi^{(\kappa)}$ displays a quasi-linear variation with $\theta$ in the LR regime, $\chi^{(p)}$ seems 
 to saturate at large $\theta$. (b) Comparison of the anomalous scaling computed in this work and found in the numerical simulations of \cite{ales2019}. 
 [The corresponding data points are courtesy of the authors].} 
\label{fig:intermexp}
\end{figure*}

  However, these conclusions do not hold if the decoupling property is not verified. To assess
   the effectiveness of decoupling,
 we have computed the non-linear part of the flow of $f_\kappa^D$ (which is related to the noise) at the fixed point $I^D_*$. The result is displayed in 
  \fig{fig:intermscal}(a), which shows  that in the LR phase $I^D_*$ becomes less and less negligible as the temporal correlation is increased.  Depending on the form of $I^D_*$ this phenomenon can modify the scaling form \eqref{eq:scalfx} in different ways. In \fig{fig:intermscal}(b) we show that in the present case the non-decoupling manifests 
 itself in the appearance of a non-zero slope in the scaling function $\zeta^D(y)$ (associated with the noise vertex) as the limit $y \rightarrow 0$ is approached, that is
\begin{equation}\label{eq:zdscal}
\zeta^D(y)=\left\{ 
\begin{matrix}
\zeta^D_0 y^{-\Delta \eta ^D} \,, & y\rightarrow 0 \\
\zeta^D_\infty y^{-\eta^D_*/z}\,, & y \rightarrow \infty
\end{matrix}
\right. \,,
\end{equation}
where $\Delta \eta^D = \eta^{D}_*-\eta^{D}_{p\gg 1}$, with $\eta^{D}_*$ the scaling exponent computed from the running
 coefficient $D_\kappa$, and   $\eta^{D}_{p\gg 1}$ the exponent computed from a power-law fit at large $p$ of the function $f^D_*$:
 $f^D_*(\vpi,p\gg 1) \sim p^{-\eta^D_{p\gg 1}}$. The behavior \eqref{eq:zdscal} corresponds to a generalized version
  of the Family-Vicsek scaling form, and is analogous to the one introduced in \cite{lopez1997} for the structure factor $S(t,\vec{k})=\langle h(t,\vec{k})h(t,-\vec{k})\rangle$.
 The value of $\eta^D_{p\gg 1}$ cannot be predicted by any scaling argument. 
 This anomalous scaling corresponds to what is termed {\it intrinsic anomalous scaling} in \cite{lopez1997}, for which $\chi<1$ but a new independent exponent exists, by opposition
 to {\it superroughnening}, for which  $\chi>1$.
This is also reminiscent of the anomalous scaling in turbulence generated by intermittency effects \cite{Frisch95}.
  We observed that in the scaling function $\zeta^\nu(y)$ the anomalous scaling is less pronounced. 
These anomalous effects increase with $\theta$, as is shown in \fig{fig:intermexp}(a) where the usual roughness exponent (denoted $\chi^{(\kappa)}$ in the following) appearing in \eqref{eq:expo} is compared with the one in which the scaling dimensions $\eta^X_*$ are replaced by the actual scaling exponents in momentum space,
\begin{equation}
\chi^{(p)}=(2-d+\eta_{p\gg 1}^D-\eta_{p\gg 1}^\nu)/2\,.
\end{equation} 
Hence the quantity $\Delta \chi =\chi^{(\kappa)} - \chi^{(p)}$ is a direct measurement of the presence of anomalous scaling in the system.

 The appearance of a new critical exponent for strong enough time correlations is  similar to the one exhibited 
 in \cite{ales2019}. In that work, this new critical exponent is linked to the presence of emergent spatial structures
  in the system, referred to as \textit{faceting}, which appears  for $\theta\gtrsim 0.23$.  In \fig{fig:intermexp}(b)
   we compare our estimate for $\Delta \chi$ with the numerical results of \cite{ales2019}. The qualitative behavior seems to be in agreement.
    Although in our case, the anomalous scaling appears in the whole LR phase $\theta>\theta_{\rm th}$,
 it becomes significant only at larger values of $\theta\gtrsim 0.3$, which is consistent with the result of \cite{ales2019}
  which reports large anomalous scaling for $\theta\gtrsim 0.3-0.4$. However, for such large values, 
 the  numerical values for $\chi$ in \cite{ales2019} are greater than one, which would correspond to superroughening, whereas
 our results remains in the regime $\chi<1$, which corresponds to intrinsic anomalous scaling.

\section{Conclusion}

In this work, we studied the effect of temporal correlations in the microscopic noise of the KPZ equation, both in $d=1$ and $d=2$. Their presence breaks the constitutive symmetry of the KPZ universality class, which is the Galilean invariance. It is thus not clear {\it a priori} whether an infinitesimal amount of temporal correlations suffice to destroy the KPZ universal properties, and this was debated in the literature.
 We investigated this issue within a non-perturbative renormalization group approach, which is functional in both momentum and frequency, and thus allows one to precisely analyze non-delta correlations in the microscopic noise.

  We first studied the case of SR temporal correlations, characterized by a finite time scale $\tau$, in $d=1$. 
 This type of correlation breaks both the Galilean and the time-reversal symmetries in $d=1$.
 However, we found that for any $\tau$, these microscopic correlations are washed out by the flow: both symmetries are dynamically restored  after a certain RG scale, and the pure SR-KPZ fixed point is reached. This means that the large distance properties of the system are still described by the KPZ class. This result is reasonable since temporal correlations 
 are hardly striclty delta-correlated in any real system and still KPZ universal properties can be observed with high accuracy in experiments \cite{Takeuchi10,Takeuchi12}.

 We then focused on the case of LR temporal correlations, embodied in a power-law with exponent $\theta$. 
In both $d=1$ and $d=2$, we found that there exists a critical value ${\theta_{\textrm{th}}}(d)$ separating two regimes, in agreement with previous RG studies in $d=1$. For $\theta<{\theta_{\textrm{th}}}$, the LR part flows to zero, the symmetries are restored and the large distance and long time universal properties are described by the pure SR-KPZ fixed point, while above this value, the LR part dominates and drives the system to a new LR fixed point, with $\theta$-dependent critical exponents $\chi(\theta)$ and $z(\theta)$ and a breaking of Galilean symmetry $\eta^\lambda_*(\theta)=z(\theta)+\chi(\theta)-2\neq 0$  increasing with $\theta$.
 We computed these exponents with increased precision compared to previous approaches in $d=1$, and provided 
 for the first time an estimate  in $d=2$.
In one dimension, we also evidenced in the LR phase some anomalous scaling. As a result, we found that the function 
associated to the noise vertex, and hence the two point correlation function, displays a scaling form which is a 
generalized version of
 the usual Family-Vicsek one, as proposed in \cite{ales2019}. This behavior may be related to the emergence of 
 intermittency effects in the system, which warrants a dedicated study. 

As a future development, it would be interesting to study the effect of temporal correlations
 within the next level of approximation, termed the SO approximation, which allows one to fully describe the momentum and frequency dependence of two-point functions without inducing any spurious breaking of symmetries.
 This will not change qualitatively the results presented here but would allow one to achieve more precision. 
 However, the flow equations at SO are much more complicated since they include contributions from all higher-order vertex functions $\Gamma_\kappa^{(n)}$ and the numerical cost to integrate them is increased. Implementing this scheme in $d>1$ would be desirable even for the pure case, since it would allow one to probe the existence of a upper critical dimension for KPZ. This is work in progress.

\begin{acknowledgments}
 The authors thank N. Wschebor and B. Delamotte for useful discussions on this work, and the authors of 
 \cite{ales2019} for kindly providing us with the data from their numerical simulations concerning the anomalous scaling.
 This work received support from the French ANR through the project NeqFluids (grant ANR-18-CE92-0019).
\end{acknowledgments}

\appendix

\section{Non-analycities in the presence of a power-law correlator}
\label{APP-power}

In principles, the presence of the regulator in the NPRG flow ensures the analyticity of all vertex functions $\Gamma_\kappa^{(n)}$ at any finite scale $\kappa$. 
 This is always true for the momentum dependence because of the presence of the regulator \eq{Rk}. However, this 
  is not guaranteed for the frequency dependence since we have not included a frequency regulator.
In fact, in most systems, as long as the initial condition is smooth, the integrands in the flow equations are generally
 well-behaved in frequency and lead to convergent integrals. A problem may arise when non-analytic initial conditions
 are considered, as for the case of LR correlations. The flow of the LR coupling $w_\kappa^\theta$ 
can be extracted from the flow of $\fd$ as
\begin{equation}
\partial_s w_\kappa^\theta=\lim_{\varpi \rightarrow 0} \varpi^{2\theta}\partial_s \fd(\varpi,0) \, .
\end{equation}
Within the NLO scheme, one finds
\begin{align}
&\partial_s w_\kappa^\theta=\lim_{\varpi\rightarrow 0}2\lambda_\kappa \nonumber\\
&\quad\quad\quad\times\int_{\omega,\vec{p}}\frac{(\vec{q}\cdot\vec{Q})^2}{P(\omega,q)^2P(\Omega,Q)}  \frac{\varpi^{2\theta}}{(\varpi+ \omega)^{2\theta}}\nonumber\\
&\times\left( P(\omega,q) \partial_s S^D_k(q)-2\vec{q}\,^2 \ell_\kappa(q) \partial_s S^\nu_k(q) \left(\frac{\varpi}{\omega}\right)^{2\theta}  \right) \, .\label{eq:dthetaflow}
\end{align}
For values $\theta<1/2$,  the contribution of the first term in the right hand side always vanishes.
 The same holds true for the second term for $\theta<1/4$. However, in the range $1/4\leq\theta<1/2$, an ambiguity arises in the second term, since this term vanishes only if the   limit $\varpi\rightarrow 0$ is taken before the integration on $\omega$.  This ambiguity is  present because the frequency sector is not properly regularized, and would disappear with a frequency-dependent regulator \cite{Duclut17}. Since the result should not depend on the choice of the regulator, we simply assume that this term is zero since it would vanish with a frequency regulator. Under this assumption, the coupling $w_\kappa^\theta$ is indeed not renormalized. The same conclusion can be reached by simply shifting the normalization point to a non-zero external frequency $\varpi_0$, which also resolves the ambiguity.

\section{Flow equation of $\lambda_{\kappa}$}\label{app:a1}

The flow of the non-linear coupling  $\lambda_{\kappa}$,  defined by \eqref{eq:fldef},  can be extracted  from the flow of the three-point function $\Gamma_\kappa^{(2,1)}$ which reads
\begin{widetext}
\begin{subequations}
{
\begin{align} \label{eq:gamma3flowso}
\partial_{\kappa} &[\GG^{(3)}_{\kappa}]_{ijk}(\pp_1,\pp_2)=\frac{1}{2}\text{Tr}\left\{ \partial_{\kappa}R_{\kappa}(\qq) G_{\kappa}(\qq) \left[ -\GG^{(5)}_{\kappa,ijk}(\pp_1,\pp_2,-\pp_1-\pp_2,\qq)\right.\right. \nonumber \\
&\left.\left.+\GG^{(4)}_{\kappa,ij}(\pp_1,\pp_2,\qq) G_{\kappa}(\pp_1+\pp_2+\qq) \GG^{(3)}_{\kappa,k}(-\pp_1-\pp_2,\pp_1+\pp_2+\qq) \right.\right.\nonumber\\
&\left.\left.-\GG^{(3)}_{\kappa,i}(\pp_1,\qq)G_{\kappa}(\pp_1+\qq)\GG^{(3)}_{\kappa,j}(\pp_2,\pp_1+\qq)G_{\kappa}(\pp_1+\pp_2+\qq)\GG^{(3)}_{\kappa,k}(-\pp_1-\pp_2,\pp_1+\pp_2+\qq)\right] G_{\kappa}(\qq)\right\}
\end{align}
}
\begin{equation}
\includegraphics[scale=1]{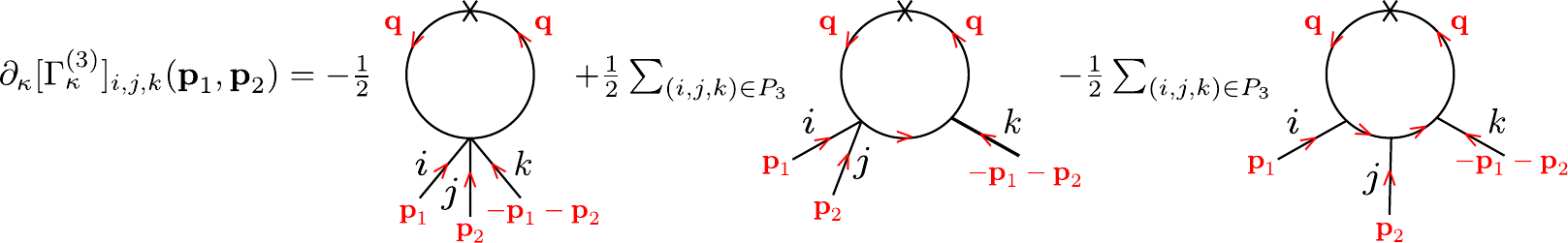}
\end{equation}
\end{subequations}
\end{widetext}
where the $\GG_\kappa^{(n)}$ on the right-hand side are represented as $2\times2$ matrices 
(\ie only the external indices are indicated, summation over the internal indices is implied),
and where the cross in the diagrams represents the derivative with respect to $\kappa$ of the regulator, and $P_n$ is the permutation group associated to the set $(i_1,\dots,i_n)$. 
 The trace operation {in \eqref{eq:gamma3flowso}} also includes all non-trivial permutations of the external vertices with associated momenta. Within the NLO$_\omega$ approximation, the 4- and 5- point vertex functions are zero, and only the 3-point vertex $\Gamma_{\kappa}^{(2,1)}\equiv [\Gamma_{\kappa}^{(3)}]_{\varphi,\varphi,\tilde{\varphi}}$ gives a non-vanishing contribution in the remaining diagrams. Evaluating this expression at external frequencies $\varpi_1=\varpi_2=0$ and external momenta $\vp_1=\vp_2\equiv\vp/2$ and   taking the limit $\vp\to 0$ yields \eq{eq:kdkflnlop}.

\section{Numerical Integration}

{In this appendix, we only consider dimensionless quantities, so we omit the hat symbols to alleviate notations.}

\label{sec:num}

\begin{figure}[b]
\includegraphics[scale=0.85]{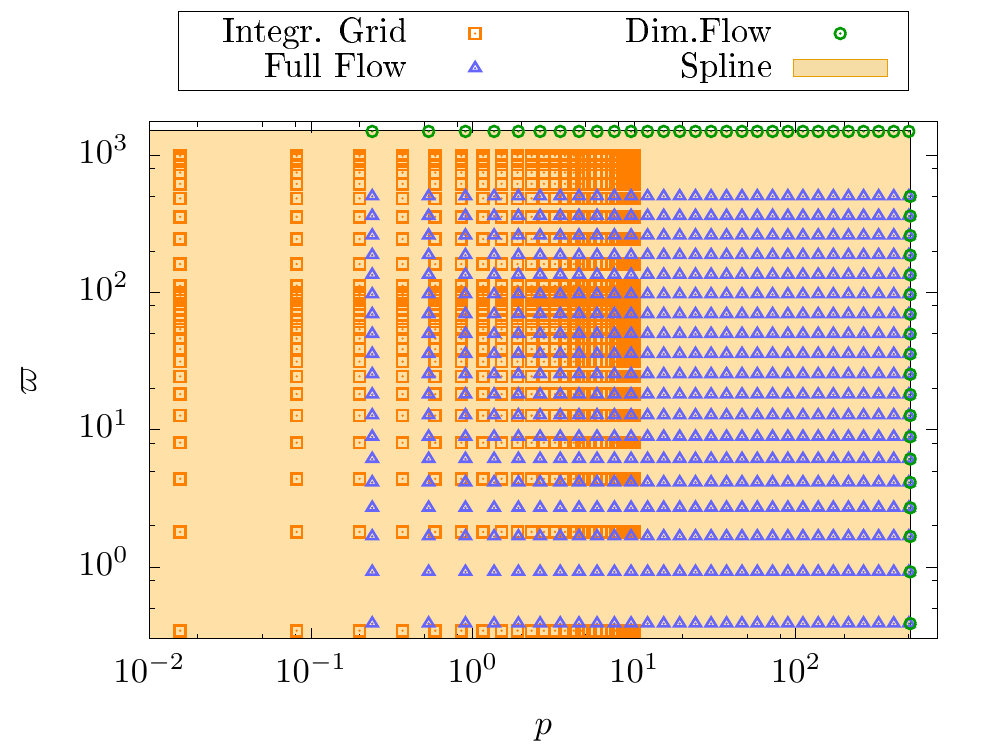}
\caption{Sketch of the numerical procedure: the blue triangles and green circles represent the grid points  where the flows of the functions $f_\kappa^X(\varpi,p)$ are computed from given initial conditions at $s=0$. The orange squares represent the grid points used in the Gauss-Legendre algorithm to compute numerically the integrals over the internal frequency and momentum (the grid for the integration over the angle is not represented in this sketch). The values of the functions in the whole orange-shaded domain are calculated using a spline interpolation. The flows of the functions on the boundaries (green circles) are approximated by   only the dimensional (linear) flow.}
\label{fig:grid}
\end{figure}

\subsection{Integration scheme}

We integrated numerically the 
 flow equations for the dimensionless functions $f_\kappa^D$, $f_\kappa^\nu$ and $f_\kappa^\lambda$, for the dimensionless couplings $g_\kappa$ and $w_\kappa^\theta$, and for the running anomalous dimensions $\eta_\kappa^\nu$ and $\eta_\kappa^D$,  using standard procedures.
The advancement in the RG time $s$
 is achieved with an adaptative time-step: 
 a default time-step $\Delta s =1$ is used as long as the ratio
 of the non-linear part $I^X_\kappa(\varpi,p)$ of the flow of a function $f^X_\kappa$ (or a coupling) for all external $(\varpi,p)$ does not exceed 1\% of the function itself $I^X_\kappa(\varpi,p)/f^{X}_\kappa(\varpi,p)<0.01$, else the time-step is iteratively decreased by a factor $\sqrt{10}$ until this constraint is satisfied. 

In generic spatial dimension $d$ there are three different integrals to perform: over the modulus of the internal momentum $q$, over the internal frequency $\omega$ and over the angle $\psi$ between the external and internal momenta.  A quadrature Gauss-Legendre method is used to compute the three of them, 
\begin{align}
\int_0^\pi d\psi \int_{0}^\infty & dq \int_{-\infty}^\infty  d\omega \, f^X_\kappa(\psi,q,\omega) \nonumber \\ 
& \rightarrow \sum_{i,j,k} w^{(\psi)}_i w^{(q)}_j w^{(\omega)}_k f^X_\kappa(\psi^*_i,q^*_j,\omega^*_k)
\end{align}
where $(\psi^*_i,q^*_j,\omega^*_k)$ are the quadrature grid-points and $w_i^{(\cdot)}$ their respective weights. For some specific points (typically zero momentum or frequency) the integrand has to be treated analytically to avoid spurious numerical divergences. The domain of integration on the internal momentum  is $q \in [0,\infty)$. However, the presence of the derivative of the regulator $\p_\kappa R_\kappa$   in $I^X_\kappa(\varpi,p)$  effectively cuts exponentially the internal momentum to $q\lesssim \kappa$ such that the integral can be performed without loss of precision over a finite domain $q \in [0,q_{max}]$. We checked that $q_{max}=10$ suffices to obtain a converged value for the integrals. On the other hand, as the frequency sector is not regularized, the integral over the internal frequency is not cut, such that the contribution of the high-frequency sector is not negligible. The integral over the internal frequency is hence performed on a domain  $\omega \in [0,\omega_{max}]$ with $\omega_{max}=10^3$.

 \begin{figure}[t]
\includegraphics[scale=0.8]{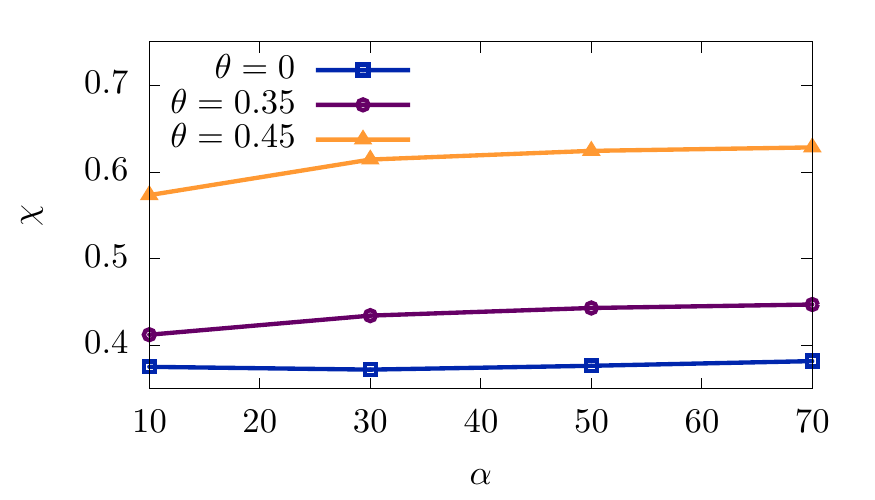}
\caption{Roughness exponent $\chi$ as a function of the cut-off parameter $\alpha$ in $d=2$ for different values of $\theta$.}
\label{fig:pms}
\end{figure}

\subsection{Grids and Interpolation}

The flow equations are computed for external frequency and momentum on grid points $(\varpi,p)$ with logarithmic
 spacing, represented by blue triangles and green circles on \fref{fig:grid}. 
To compute the integrals over the internal momentum $q$ and frequency $\omega$, the functions $f_\kappa^X$ have to be evaluated  at values $Q=|\vec{p}+\vec{q}|$ and $\Omega=\omega+\varpi$, which can fall outside grid points. In this work, these integrals are computed using another grid, represented in  \fref{fig:grid} by orange squares, which corresponds to the Gauss-Legendre quadrature roots in their respective integration domains, $q \in [0,q_{max}]$ and $\omega \in [0,\omega_{max}]$. 
The external momenta (respectively frequencies) are chosen such that 
$p\in [0,Q_{max}=p_{max}+q_{max}]$ (resp. $\varpi\in [0,\Omega_{max}=\varpi_{max}+\omega_{max}]$), where $p_{max}$ (resp. $\varpi_{max})$) is the last blue triangle and $Q_{max}$ (resp. $\Omega_{max}$) is the following green circle. 
The functions can be  evaluated in the whole orange-shaded domain using a bi-cubic spline procedure from the external points (represented by blue triangles and green circles). The values of the derivatives $\p_p {f}^X_\kappa(\varpi,p)$ and $\p_\varpi {f}^X_\kappa(\varpi,p)$ are also evaluated using  the bi-cubic spline interpolation. With this choice, the flow equations of ${f}^X_\kappa(\varpi,p)$ for all grid points up to  $(\varpi_{max},p_{max})$ can  hence be evaluated since the values of integration all lie in the orange-shaded domain. To compute the flow on the boundaries, \ie the points with $\varpi= \Omega_{max}$ or $p=Q_{max}$, represented by the green circles
 on \fref{fig:grid}, we exploit the decoupling property of the flow: for sufficiently high momenta and frequencies,
 the non-linear part of the flow $I^X_\kappa(\varpi,p)$ become negligible compared to the function ${f}^X_\kappa$ itself when the fixed-point is approached. For these points, we thus approximate the flow by the linear contribution only:
\begin{equation}
\p_s {f}^X_\kappa(\varpi,p)=\bigg(\eta^X_\kappa + (2-\eta^\nu_\kappa){\varpi} \p_{{\varpi}} + {p}\,\p_{{p}}\bigg) {f}^X_\kappa(\varpi,p).
\end{equation}

\subsection{Choice of the cutoff parameter and normalization point}
\label{sec:cutoff}

The cutoff function \eq{eq:expReg} depends on a free parameter $\alpha$, which can be varied to assess 
the precision within  a chosen approximation level.
 Indeed, if the flow equations were solved exaclty, the results would not depend on the regulator. Any approximation 
 induces a spurious residual dependence on $\alpha$, and a minimal sensitivity principle can be exploited to select the optimal value of $\alpha$ \cite{Canet04a}. 
 In both $d=1$ and $d=2$,  we studied the influence of $\alpha$ on the critical exponents  obtained at both NLO and NLO$_\omega$. We observed that their values only weakly depend on  $\alpha$,  as illustrated for $\chi$ in \fref{fig:pms}. We thus fixed $\alpha=10$ for all the results presented in this work. This value corresponds to an optimal value for the pure KPZ case in $d=1$ at NLO \cite{Kloss12}.

\begin{figure}[b]
\includegraphics[scale=0.8]{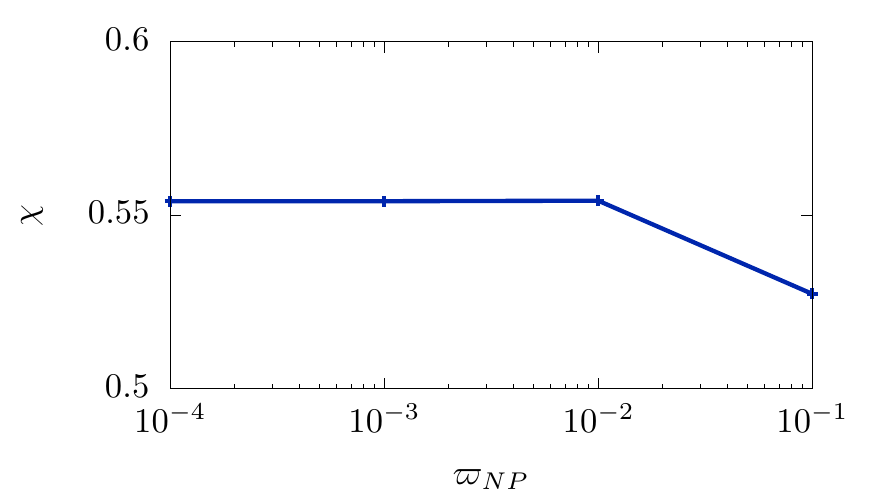}
\caption{Roughness exponent $\chi$ for $\theta=0.45$ as a function of the normalization point $\varpi_0$ in $d=2$.}
\label{fig:pmsnp}
\end{figure}

Let us finally discuss the choice of the normalization point $\varpi_0$ in $d=2$.
{The normalization point corresponds to a specific configuration of the external momentum and frequency.
 While the RG flow equations for the running parameters $X_\kappa$ depend on the particular configuration chosen 
 to define it $X_\kappa=f^X_\kappa(\varpi_{\rm NP},p_{\rm NP})$,
  the running dimensions $\eta^X_\kappa$ at the fixed point do not \cite{zinnjustin2002,tauber2014}. 
  It is interesting to note that in the framework of NPRG, one can in general safely take the simplest choice
   $(\varpi_{\rm NP},p_{\rm NP})=(\varpi=0,p=0)$, which is a problematic configuration in the  dimensional-regularization
    scheme due to the mixing of infra-red and ultra-violet singularities for massless theories \cite{frey1994}.
    
     In this work, as the power-law exponent $\theta$ for the time correlations approaches the value $\theta=1/4$, 
     a divergence emerges in the flow of $f^D_\kappa(\varpi=0,p)$. As mentioned in the main text, this divergence is a 
     pure artifact due to the absence of a frequency regulator. As explained in \sref{sec:d2}, one can simply avoid such a 
     singularity by shifting the normalization point to a small but non-zero value $\varpi_0$. 
     As we are working in practice with some approximations and not with the exact theory,
      varying the normalization point introduces a small spurious variation of the anomalous  dimensions
       $\eta^X_*$ at the fixed point. However, we checked that for small enough $\varpi_0$, this residual
        dependence is indeed negligible. This is illustrated on \fref{fig:pmsnp} which shows the variation
         of $\chi$ with $\varpi_0$ for $\theta=0.45$, for which the singularity at the origin is the steepest.
          We hence fixed $\varpi_0=0.01$ in $d=2$ for all values of $\theta$.


\end{document}